\documentclass[11pt]{article}

\usepackage{sn-preamble}
\usepackage{soul}
\usepackage{microtype}
\usepackage{caption}
\usepackage{float}
\usepackage{needspace}

\definecolor{ovInk}{HTML}{26364A}
\definecolor{ovMuted}{HTML}{637181}
\definecolor{ovLine}{HTML}{A3A3A3}
\definecolor{ovArrow}{HTML}{A3A3A3}
\definecolor{ovHead}{HTML}{3C5E91}
\definecolor{ovTail}{HTML}{B4772C}
\definecolor{ovLocal}{HTML}{287F79}
\tikzset{
    ov head/.style={draw=ovHead,fill=ovHead!35,line width=.75pt},
    ov tail/.style={draw=ovTail,fill=ovTail!20,line width=.75pt},
    ov arrow/.style={-latex,draw=ovInk!85,line width=.85pt},
    ov connection/.style={ov arrow,draw=ovArrow},
    ov vertical connection/.style={ov connection,shorten <=2mm,shorten >=2mm},
    ov label/.style={inner sep=0pt,font=\rmfamily\small},
    ov equation/.style={inner sep=0pt,font=\rmfamily\normalsize},
    ov tool/.style={fill=white,inner xsep=5pt,inner ysep=3pt,
        align=center,font=\rmfamily\footnotesize}
}

\crefname{appendix}{appendix}{appendices}
\Crefname{appendix}{Appendix}{Appendices}

\DeclareMathOperator{\Tr}{Tr}

\DeclareMathOperator{\spn}{span}

\DeclareMathOperator{\corr}{corr}
\newcommand{\OPT}{\mathrm{OPT}}
\newcommand{\Eigen}{\mathrm{Eigen}}
\newcommand{\err}{\mathrm{err}}

\newcommand{\G}{{\mathrm{G}}}

\title{Efficient Proper Agnostic Learning of Halfspaces}
\author{
    Anindya De\thanks{University of Pennsylvania.
    Email:~\href{mailto:anindyad@seas.upenn.edu}{\texttt{anindyad@seas.upenn.edu}}.}
    \and
    Elchanan Mossel\thanks{MIT.
    Email:~\href{mailto:elmos@mit.edu}{\texttt{elmos@mit.edu}}.}
    \and
    Shivam Nadimpalli\thanks{MIT.
    Email:~\href{mailto:shivamn@mit.edu}{\texttt{shivamn@mit.edu}}.}
    \and
    Joe Neeman\thanks{
    Email:~\href{mailto:joeneeman@gmail.com}{\texttt{joeneeman@gmail.com}}.}
}
\date{\today}

\begin{document}

\hypersetup{pageanchor=false}
\pagenumbering{gobble}

\maketitle

\begin{abstract}
We give a \emph{proper} membership-query algorithm for agnostically learning halfspaces under the uniform distribution on $\{\pm1\}^n$.
Given query access to an arbitrary function $f:\{\pm1\}^n \to \{\pm1\}$, our algorithm runs in time $\mathrm{poly}(n)+2^{\mathrm{poly}(1/\varepsilon)}$ and outputs a halfspace with error at most $\mathrm{OPT}+\varepsilon$, where $\mathrm{OPT}$ is the error of the best-fitting halfspace for $f$.
Previously, even with membership queries, the best known running time was $n^{\mathrm{poly}(1/\varepsilon)}$, achieved by the sample-based algorithm of Kalai, Klivans, Mansour, and Servedio (FOCS~2005).
Our algorithm is the first agnostic halfspace learner under the uniform distribution on $\{\pm1\}^n$ whose running time is polynomial in $n$ with an exponent independent of $\varepsilon$.

Combining our algorithm with a boosting argument yields a proper agnostic learner using only random examples.
It achieves the same $\mathrm{OPT}+\varepsilon$ guarantee, with running time matching that of Kalai et al.\ up to an additive term depending only on $\varepsilon$.
To our knowledge, this is the first sample-based proper agnostic learner in the Boolean setting with running time $2^{o(n)}$ for every fixed $\varepsilon>0$.
\end{abstract}

\newpage

\tableofcontents
\newpage

\setcounter{page}{1}
\pagenumbering{arabic}
\hypersetup{pageanchor=true}

\section{Introduction}
\label{sec:introduction}

Halfspaces, also known as linear threshold functions, are among the most basic objects in learning theory.
Their simple geometric form makes them readily interpretable, and they have been a central topic of study since the earliest days of machine learning~\cite{Rosenblatt:58,MinskyPapert:68}. 
They also underlie widely used classification methods, including support vector machines and kernel-based classifiers, which learn linear separators in suitable feature spaces~\cite{BGV:92,CortesVapnik:95}.
In Valiant's PAC model~\cite{Valiant:84}, halfspaces are efficiently learnable under \emph{any input distribution} in the realizable setting: given examples labeled by an unknown halfspace, one can find a consistent halfspace using linear programming~\cite{BEH+:89}.

The problem becomes substantially more challenging in the presence of noise, captured by the \emph{agnostic learning} model of Kearns, Schapire, and Sellie~\cite{KSS:94}.
In this setting, the labels may be arbitrary, and the learner seeks a hypothesis whose error is within an additive $\eps$ of that of the best-fitting halfspace.
For arbitrary input distributions, however, strong computational hardness results obstruct efficient learning.
These include hardness of finding a good halfspace hypothesis~\cite{GuruRag09} and, under widely studied complexity assumptions, hardness even when the learner may output an arbitrary classifier~\cite{Daniely16,Tiegel23}. 
These barriers have motivated extensive work on \emph{distribution-specific} learning, where assumptions on the input distribution make progress possible while leaving the labels unrestricted.
This approach has yielded agnostic learners under the uniform Boolean and Gaussian distributions~\cite{KKMS:08,DKKTZ21}, as well as more general log-concave distributions~\cite{KKM13}.

Our focus is on agnostic learning under the uniform distribution on the Boolean hypercube $\bn$. 
Let $\calH_n$ denote the class of affine halfspaces $h:\bits^n\to\bits$ of the form $h(x)=\sign(w\cdot x-\theta)$, where $w\in\R^n$ and $\theta\in\R$.
For an arbitrary target function $f:\bits^n\to\bits$ and a hypothesis $g:\bits^n\to\bits$, define
\[
    \err_f(g):=\Prx_{\bx\sim\bits^n}[f(\bx)\ne g(\bx)]\,,
    \qquad
    \OPT_f=\OPT_f(\calH_n):=\min_{h\in\calH_n}\err_f(h)\,.
\]
Given access to $f$, the goal is to output a hypothesis $g$ satisfying $\err_f(g)\le\OPT_f+\eps$.
Thus, the learner's output must compete with the best-fitting halfspace up to an arbitrarily small additive error. 
Without the computational aspects,
this question in fact predates the agnostic learning model and traces back to Vapnik's early work on empirical risk minimization~\cite{Vap82,Vapnik:98}.
 
In a landmark result, Kalai, Klivans, Mansour, and Servedio~\cite{KKMS:08} gave an agnostic learner for this problem with running time $n^{O(\eps^{-2}\log(1/\eps))}$.\footnote{The original bound obtained by~\cite{KKMS:08} was $n^{O(1/\eps^4)}$.
The stated bound follows from the improved $L^1$ polynomial approximation of Feldman, Kothari, and Vondr\'ak~\cite[Corollary~15]{FKV17}.}  
The algorithm uses only uniformly random labeled examples $(\bx,f(\bx))$ and is remarkably simple: it fits a low-degree polynomial to uniformly random labeled examples using $L^1$ regression, then thresholds the resulting polynomial.
Kalai et al.~\cite[Section~4.2]{KKMS:08} also showed that a sample-based agnostic learner with total running time \smash{$n^{O(\eps^{-2+\delta})}$} for some fixed $\delta > 0$ would yield a substantially faster algorithm for the notoriously difficult problem of learning parities with noise.
This suggests that major improvements in the dependence on the dimension may require going beyond random examples. 

For over two decades, the polynomial-regression approach of Kalai et al.~\cite{KKMS:08} has remained the best-known sample-based agnostic learner for halfspaces under the uniform Boolean distribution.
Faster algorithms have been developed with relaxed ``quasi-agnostic'' guarantees (see~\Cref{subsec:related-work}), but these do not achieve the additive guarantee $\OPT_f+\eps$.
The best known approach therefore retains two limitations: it is \emph{improper}, returning a polynomial threshold function rather than a halfspace, and greater accuracy comes at the cost of a worse exponent in the dimension. 

\subsection{Our Results}
\label{subsec:our-results}

The hardness result above motivates seeking improvements over~\cite{KKMS:08} through stronger access to the target function. 
A natural model allows \emph{membership queries}, in which the learner evaluates $f$ on inputs of its choosing.
Feldman~\cite{Feldman09MQ} explicitly asked whether such access can offer a computational advantage in distribution-specific agnostic learning.
He established such an advantage for a particular concept class under cryptographic assumptions, while leaving open whether it is possible for natural classes such as halfspaces.
Gopalan, Kalai, and Klivans~\cite{GKK08open} posed the stronger quantitative question of whether membership queries permit agnostic learning of halfspaces under the uniform Boolean distribution in time $\poly(n,1/\eps)$.

We answer Feldman's question affirmatively for halfspaces, giving the first improvement over polynomial regression~\cite{KKMS:08} in the Boolean setting. 
Our learner uses $\poly(n,1/\eps)$ queries and is moreover \emph{proper}: its output is itself a halfspace.

\begin{restatable}[Query-based proper agnostic learner]{theorem}{MainTheorem}
	\label{thm:main}
	Given
	$\eps\in(0,1/2)$ and query access to
	$f:\bits^n\to\bits$, there is an algorithm
	(\Cref{alg:global-agnostic-learner}) that outputs a halfspace $g\in\calH_n$ satisfying
	\[
		\err_f(g)\le\OPT_f+\eps\,,
		\]
		with probability at least $2/3$.
		It makes at most
		\smash{$O(n\eps^{-28})$}
		queries to $f$ and runs in time at most
		\smash{$O(n^3\eps^{-30})+2^{O(\eps^{-14}\log(1/\eps))}$}.
\end{restatable}

All running-time bounds in this paper count arithmetic operations in the real-RAM model. 
The algorithm underlying \Cref{thm:main} and its guarantees also extend to bounded real-valued target functions; see \Cref{rem:bounded-real-targets}.  

A comparable running-time guarantee in Gaussian space follows from the work of Diakonikolas, Kane, Kontonis, Tzamos, and Zarifis~\cite{DKKTZ24}, and was implicit in the work of De, Mossel, and Neeman~\cite{DMN21}. 
We discuss these connections and the challenges of extending the Gaussian techniques to the Boolean setting in \Cref{subsec:overview-main,subsec:related-work}. 
To our knowledge, ours is the first agnostic learner for halfspaces on the Boolean hypercube with running time polynomial in $n$ and an exponent independent of $\eps$. 

Combining our query-based learner with a boosting argument~\cite{GHKRW23}, we also obtain an efficient \emph{proper} agnostic learner from random examples alone. 
To state this result, let $D$ be a distribution on $\bits^n\times\bits$
whose marginal on $\bx$ is uniform, and write
\[
    \err_D(h):=\Prx_{(\bx,\by)\sim D}[h(\bx)\ne\by]\,,
    \qquad
    \OPT_D:=\min_{h\in\calH_n}\err_D(h)\,.
\]
Here the labels may be randomized even after conditioning on $\bx$.

\begin{restatable}[Sample-based proper agnostic learner]{theorem}{ProperSampleTheorem}
	\label{thm:proper-sample}
	Let $D$ be any distribution on $\bits^n\times\bits$ whose marginal on
	$\bits^n$ is uniform.
	For every $\eps\in(0,1/2)$, there is an algorithm using only independent labeled examples from $D$ that, with probability at least $2/3$, outputs a halfspace $g\in\calH_n$ satisfying
	\[
		\err_D(g)\le\OPT_D+\eps\,.
	\]
	It uses \smash{$n^{O(\eps^{-2}\log(1/\eps))}\poly(1/\eps)$} labeled examples and runs in time at most
	$\smash{n^{O(\eps^{-2}\log(1/\eps))}\poly(1/\eps)}+\allowbreak\smash{2^{O(\eps^{-14}\log(1/\eps))}}$.
\end{restatable}

\Cref{thm:proper-sample} thus achieves the additive guarantee of~\cite{KKMS:08} with the same asymptotic dependence on $n$, while returning a halfspace. 
To our knowledge, even a $2^{o(n)}$-time proper agnostic learner achieving error $\OPT+\eps$ for every fixed $\eps>0$ was not previously known under the uniform distribution over $\bn$.   
The additive guarantee is essential here: efficient proper learners with weaker quasi-agnostic guarantees were already known (see~\Cref{subsec:related-work}). 
We also note that an analogue of~\Cref{thm:proper-sample} was known in the Gaussian setting due to Diakonikolas et al.~\cite{DKKTZ21}. 

Note that the conclusion of \Cref{thm:proper-sample} does not follow by running the \cite{KKMS:08} improper learner once and then finding a halfspace close to its output using~\Cref{thm:main}. 
Even if one could solve the latter problem efficiently, the usual triangle-inequality
argument gives only a $3\OPT+O(\eps)$ guarantee; see the discussion in
\Cref{sec:proper-sample-learning}. 

\subsection{\texorpdfstring{Overview of \Cref{thm:main}}{Overview of Theorem 1}}
\label{subsec:overview-main}

We now give an overview of the query-based agnostic learner underlying \Cref{thm:main}. 
We assume familiarity with basic Fourier analysis on the Boolean cube and Gaussian space; see~\Cref{sec:preliminaries} for a brief review. 

At a high level, the algorithm performs dimension reduction followed by empirical risk minimization.
Given an unknown target function $f\isafunc$, we identify a low-dimensional space of linear features such that restricting our search to halfspaces over these features increases the optimal error by at most $\eps$.
This strategy also appears in work on testing low-dimensional functions~\cite{DMN19,DMN21,nadimpalli2024optimal} and in agnostic learning under the Gaussian distribution~\cite{DKKTZ21,DKKTZ24}.
The main challenge is to identify the right feature space from the unknown target $f$.
To do so, we develop a Boolean analogue of the ``influence PCA'' machinery used in the Gaussian setting~\cite{DMN21,DKKTZ21,DKKTZ24}; this is our main conceptual contribution and may be of independent interest.

Fix an arbitrary halfspace $h(x)=\sign(w\cdot x-\theta)$ over $\bn$.
It is convenient to work with correlation rather than error:
\[
    \corr_f(h)
    :=
    1-2\err_f(h)
    =
    \Ex_{\bx\sim\bn}\sbra{f(\bx)\,h(\bx)}\,.
\]
We will show that $h$ can be replaced, with little loss in correlation, by a halfspace whose normal lies in a low-dimensional subspace determined only by $f$.
The same subspace will therefore contain a nearly optimal competitor, regardless of the original halfspace $h$. 

\paragraph{The Boolean Influence PCA Matrix.}

To motivate our construction, we recall how the Gaussian approach~\cite{DMN21,DKKTZ21,DKKTZ24} uses gradients to identify relevant directions. 
One smooths the target to obtain a differentiable function $F:\R^n\to[-1,1]$, then forms the gradient second-moment matrix 
\[
    \Gamma(F)
    :=
    \Ex_{\bx\sim N(0,I_n)}
    \sbra{\nabla F(\bx)\,\nabla F(\bx)^\top}\,.
\]
With suitable smoothing, the eigenvectors whose eigenvalues exceed an appropriate threshold span a subspace of dimension independent of $n$ that contains the normal vector of a nearly optimal halfspace. 

On the Boolean cube, directly taking the second moment of the un-smoothed discrete gradient can be uninformative.
Writing
\[
    \D_i f(x)
    :=
    \frac{f(x^{i\mapsto+1})-f(x^{i\mapsto-1})}{2}\,,
    \qquad
    \D f(x):=(\D_1f(x),\ldots,\D_nf(x))\,,
\]
consider the matrix
\[
    A(f)
    :=
    \Ex_{\bx\sim\bn}
    \sbra{\D f(\bx)\,\D f(\bx)^\top}\,.
\]
The high-frequency components of $f$ can make this matrix spectrally uninformative, even when $f$ correlates well with a halfspace. 
For example, for $n\ge3$, consider
\[
    f(x)
    =
    \begin{cases}
        x_2 & \text{if }x_1=+1\\
        \prod_{i=3}^n x_i & \text{if }x_1=-1
    \end{cases}\,.
\]
A direct calculation gives $A(f)=\tfrac12 I_n$, although the dictator $x_2$ has correlation $1/2$ with $f$.
Thus, eigenvalue thresholding retains either all $n$ directions or none.

Averaging the discrete gradient first loses too much information in the other direction: it retains only the degree-$1$ Fourier coefficients, since
\[
    \Ex_{\bx\sim\bn}\sbra{\D_i f(\bx)}=\wh f(i)\,.
\]
These coefficients can vanish even when $f$ has substantial correlation with a halfspace.
For example, $f(x)=x_1x_2x_3$ has no degree-$1$ Fourier coefficients, yet $-\sign(x_1+x_2+x_3)$ agrees with it on a $3/4$ fraction of the cube.

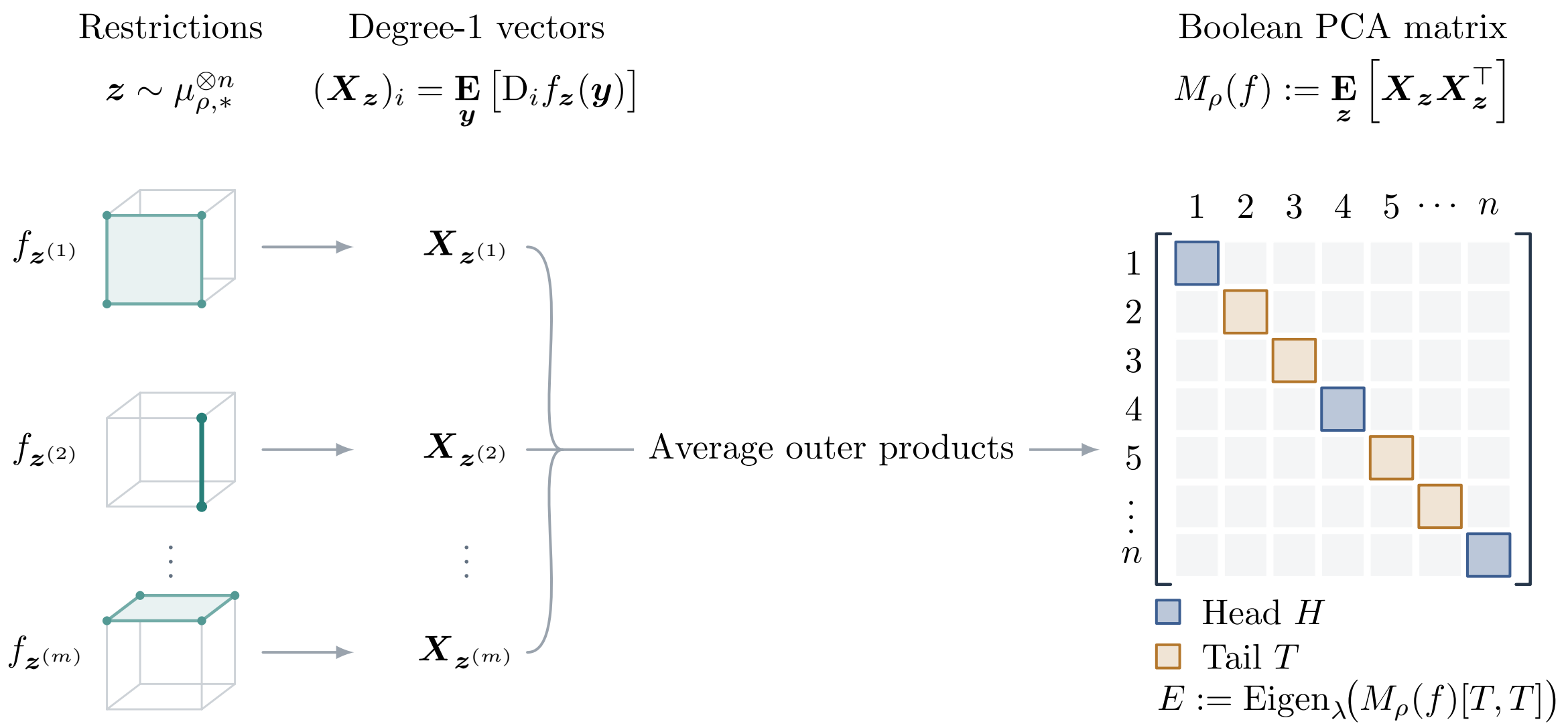
\begin{figure}[t]
    \centering
\begingroup
\tikzset{
  qc arrow/.style={-latex,draw=ovMuted!65,line width=.75pt},
  qc cube/.pic={
    \coordinate (a) at (0,0);
    \coordinate (b) at (.936,0);
    \coordinate (c) at (.936,.8736);
    \coordinate (d) at (0,.8736);
    \coordinate (e) at (.3276,.2496);
    \coordinate (f) at (1.2636,.2496);
    \coordinate (g) at (1.2636,1.1232);
    \coordinate (h) at (.3276,1.1232);
    \draw[ovMuted!32,line width=.55pt]
      (a)--(b)--(c)--(d)--cycle
      (e)--(f)--(g)--(h)--cycle
      (a)--(e) (b)--(f) (c)--(g) (d)--(h);
  }
}
\begin{tikzpicture}[x=1cm,y=1cm,text=black,font=\rmfamily\small,
  every node/.style={inner sep=0pt},line cap=round,line join=round]
\path[use as bounding box] (-8,-1.55) rectangle (8,-8.80);

\node[anchor=base] at (-6.17,-1.94) {Restrictions};
\node[anchor=base] at (-6.17,-2.55) {$\bz\sim\mu_{\rho,*}^{\otimes n}$};
\node[anchor=base] at (-3.15,-1.94) {Degree-$1$ vectors};
\node[anchor=base] at (-3.15,-2.55)
  {$\displaystyle (\bX_{\bz})_i=\Ex_{\by}\sbra{\D_i f_{\bz}(\by)}$};
\node[anchor=base] at (5.40,-1.94) {Boolean PCA matrix};
\node[anchor=base] at (5.40,-2.55)
  {$\displaystyle M_\rho(f)\mathrel{\ordinarycolon=}
    \Ex_{\bz}\sbra{\bX_{\bz}\bX_{\bz}^\top}$};

\begin{scope}[yshift=-.4cm]
\begin{scope}[shift={(-6.80,-4.162)}]
  \pic {qc cube};
  \path[fill=ovLocal!10,draw=ovLocal!65,line width=.85pt]
    (a)--(b)--(c)--(d)--cycle;
  \foreach \p in {a,b,c,d} \fill[ovLocal!80] (\p) circle (1.25pt);
\end{scope}
\node at (-7.40,-3.60) {$f_{\bz^{(1)}}$};
\draw[qc arrow] (-5.26,-3.60)--(-4.36,-3.60);
\node at (-3.25,-3.60) {$\bX_{\bz^{(1)}}$};

\begin{scope}[shift={(-6.80,-6.162)}]
  \pic {qc cube};
  \draw[ovLocal,line width=1.35pt] (b)--(c);
  \foreach \p in {b,c} \fill[ovLocal] (\p) circle (1.5pt);
\end{scope}
\node at (-7.40,-5.60) {$f_{\bz^{(2)}}$};
\draw[qc arrow] (-5.26,-5.60)--(-4.36,-5.60);
\node at (-3.25,-5.60) {$\bX_{\bz^{(2)}}$};

\node[text=ovMuted] at (-6.17,-6.60) {$\vdots$};
\node[text=ovMuted] at (-3.25,-6.60) {$\vdots$};

\begin{scope}[shift={(-6.80,-8.162)}]
  \pic {qc cube};
  \path[fill=ovLocal!10,draw=ovLocal!65,line width=.85pt]
    (d)--(c)--(g)--(h)--cycle;
  \foreach \p in {d,c,g,h} \fill[ovLocal!80] (\p) circle (1.25pt);
\end{scope}
\node at (-7.40,-7.60) {$f_{\bz^{(m)}}$};
\draw[qc arrow] (-5.26,-7.60)--(-4.36,-7.60);
\node at (-3.25,-7.60) {$\bX_{\bz^{(m)}}$};

\draw[ovMuted!65,line width=.75pt] (-2.65,-3.60)
  .. controls (-2.15,-3.60) and (-2.65,-5.60) .. (-2.30,-5.60);
\draw[ovMuted!65,line width=.75pt] (-2.65,-5.60)--(-2.30,-5.60);
\draw[ovMuted!65,line width=.75pt] (-2.65,-7.60)
  .. controls (-2.15,-7.60) and (-2.65,-5.60) .. (-2.30,-5.60);
\draw[qc arrow] (-2.30,-5.60)
  --node[midway,fill=white,inner xsep=4pt,inner ysep=3pt]
    {Average outer products}
    (3.00,-5.60);
\end{scope}

\begin{scope}[shift={(3.75,-3.95)}]
  \foreach \r in {0,...,6} {
    \foreach \c in {0,...,6} {
      \path[fill=ovInk!5,draw=white,line width=.3pt]
        ({.48*\c},{-.48*\r}) rectangle ++(.42,-.42);
    }
  }
  \foreach \i in {0,...,6}
    \path[fill=ovInk!11]
      ({.48*\i},{-.48*\i}) rectangle ++(.42,-.42);
  \foreach \i in {0,3,6}
    \path[ov head] ({.48*\i},{-.48*\i}) rectangle ++(.42,-.42);
  \foreach \i in {1,2,4,5}
    \path[ov tail] ({.48*\i},{-.48*\i}) rectangle ++(.42,-.42);
  \foreach \i/\idx in {0/1,1/2,2/3,3/4,4/5,6/n} {
    \node at ({.48*\i+.21},.35) {$\idx$};
    \node[anchor=east] at (-.32,{-.48*\i-.21}) {$\idx$};
  }
  \node at (2.61,.35) {$\cdots$};
  \node at (-.44,-2.61) {$\vdots$};
  \draw[ovInk,line width=.85pt]
    (-.05,.08)--(-.19,.08)--(-.19,-3.38)--(-.05,-3.38);
  \draw[ovInk,line width=.85pt]
    (3.36,.08)--(3.50,.08)--(3.50,-3.38)--(3.36,-3.38);
\end{scope}
\path[ov head] (3.56,-7.485) rectangle ++(.23,-.23);
\node[anchor=west] at (4.00,-7.60) {Head $H$};
\path[ov tail] (3.56,-7.925) rectangle ++(.23,-.23);
\node[anchor=west] at (4.00,-8.04) {Tail $T$};
\node[anchor=west] at (3.56,-8.48)
  {$E\mathrel{\ordinarycolon=}\mathrm{Eigen}_{\lambda}\!\bigl(M_\rho(f)[T,T]\bigr)$};
\end{tikzpicture}
\endgroup
    \vspace{1em}
    \caption{%
    \Cref{alg:global-agnostic-learner} uses membership queries to estimate degree-$1$ Fourier vectors of random restrictions of $f$, equivalently their average discrete gradients. 
    The Boolean influence PCA matrix $M_\rho(f)$ is the expected outer product of these vectors. 
    It determines the head variables $H$, the tail variables $T$, and the leading tail eigenspace $E$. 
    }
    \label{fig:query-construction}
\end{figure}

\begin{figure}[!t]
    \centering
\begin{tikzpicture}[
    x=1cm,y=1cm,font=\rmfamily\normalsize,text=ovInk,
    line cap=round,line join=round,
    ct axis/.style={draw=ovMuted!85,line width=.6pt},
    ct divider/.style={draw=ovMuted!80,densely dashed,line width=.7pt},
    ct step/.style={anchor=east,inner sep=0pt,font=\rmfamily\mdseries\small},
    ov label/.append style={font=\normalfont\normalsize},
    ov equation/.append style={font=\normalfont\normalsize},
    ov vertical connection/.append style={draw=ovMuted!85,line width=.9pt}
]

\foreach \step in {0,1,2,3}
    \node[ct step] at (-6.25,{.65-4.50*\step}) {Step \step:};

\begin{scope}[shift={(-4.11,0)}]
    \foreach \i/\h in {1/1.13,5/.54,9/1.39,12/.87}
        \path[ov head] ({.56*\i},0) rectangle ++(.38,\h);
    \foreach \i/\h in {0/.31,2/.42,3/1.62,4/.27,6/.39,7/.30,8/1.23,10/.43,11/.34,13/.26,14/.38}
        \path[ov tail] ({.56*\i},0) rectangle ++(.38,\h);
    \draw[ct axis] (-.08,0)--(8.30,0);
    \node[ov label] at (4.11,-.27) {$w$};
\end{scope}
\node[ov equation] at (0,-.68) {$h(x)=\sign(w\cdot x-\theta)$};

\draw[ov vertical connection] (0,-.98)--node[midway,ov tool]
    {Split into head and tail variables}
    (0,-2.72);

\begin{scope}[shift={(-4.11,-4.50)}]
    \foreach \i/\h in {0/1.13,1/.54,2/1.39,3/.87}
        \path[ov head] ({.49*\i},0) rectangle ++(.38,\h);
    \foreach \i/\h in {0/.31,1/.42,2/1.62,3/.27,4/.39,5/.30,6/1.23,7/.43,8/.34,9/.26,10/.38}
        \path[ov tail] ({2.70+.51*\i},0) rectangle ++(.38,\h);
    \draw[ct axis] (-.08,0)--(8.30,0);
    \draw[ct divider] (2.26,-.09)--(2.26,1.71);
    \node[ov label,text=ovHead] at (.92,-.27) {$w_H$};
    \node[ov label,text=ovTail] at (5.44,-.27) {$w_T$};
\end{scope}
\node[ov equation] at (0,-5.18)
    {$h(x)=\sign\bigl({\color{ovHead}w_H\cdot x_H}
        +{\color{ovTail}w_T\cdot x_T}-\theta\bigr)$};

\draw[ov vertical connection] (0,-5.48)--node[midway,ov tool]
    {Critical-index dichotomy to obtain a regular tail (\Cref{lem:global-regular-tail})}
    (0,-7.22);

\begin{scope}[shift={(-4.11,-9.00)}]
    \foreach \i/\h in {0/1.13,1/.54,2/1.39,3/.87}
        \path[ov head] ({.49*\i},0) rectangle ++(.38,\h);
    \foreach \i/\h in {0/.31,1/.42,3/.27,4/.39,5/.30,7/.43,8/.34,9/.26,10/.38}
        \path[ov tail] ({2.70+.51*\i},0) rectangle ++(.38,\h);
    \foreach \i in {2,6}
        \draw[ovTail,densely dashed,line width=.65pt]
            ({2.70+.51*\i},.025) rectangle ++(.38,.11);
    \draw[ct axis] (-.08,0)--(8.30,0);
    \draw[ct divider] (2.26,-.09)--(2.26,1.71);
    \node[ov label,text=ovHead] at (.92,-.27) {$w_H$};
    \node[ov label,text=ovTail] at (5.44,-.27) {$u$};
\end{scope}
\node[ov equation] at (0,-9.68)
    {$h_{\mathrm{reg}}(x)=\sign\bigl({\color{ovHead}w_H\cdot x_H}
        +{\color{ovTail}u\cdot x_T}-\theta'\bigr)$};

\draw[ov vertical connection] (0,-9.98)--node[midway,ov tool]
    {Closeness of halfspace extensions (\Cref{thm:boolean-gaussian-bridge})
     \& Poincar\'e inequality (\Cref{lem:poincare-projection})}
    (0,-11.72);

\begin{scope}[shift={(-4.11,-13.50)}]
    \foreach \i/\h in {0/1.13,1/.54,2/1.39,3/.87}
        \path[ov head] ({.49*\i},0) rectangle ++(.38,\h);
    \foreach \i/\h in {0/.32,1/.46,2/.37,3/.43,4/.29}
        \path[ov tail] ({2.70+.51*\i},0) rectangle ++(.38,\h);
    \draw[ct axis] (-.08,0)--(8.30,0);
    \draw[ct divider] (2.26,-.09)--(2.26,1.71);
    \node[ov label,text=ovHead] at (.92,-.27) {$w_H$};
    \node[ov label,text=ovTail] at (5.44,-.27) {$\Pi_Eu$};
\end{scope}
\node[ov equation] at (0,-14.18)
    {$h'(x)=\sign\bigl({\color{ovHead}w_H\cdot x_H}
        +{\color{ovTail}(\Pi_Eu)\cdot x_T}-\theta''\bigr)$};
\end{tikzpicture}
    \caption{%
    Transforming a candidate halfspace into one with tail normal in $E$, with little loss in correlation.
    The head weights remain unchanged, while the threshold may change.
    The final tail (Step~3) is depicted schematically and need not be sparse in the original coordinate basis.
    }
    \label{fig:candidate-halfspace-overview}
\end{figure}
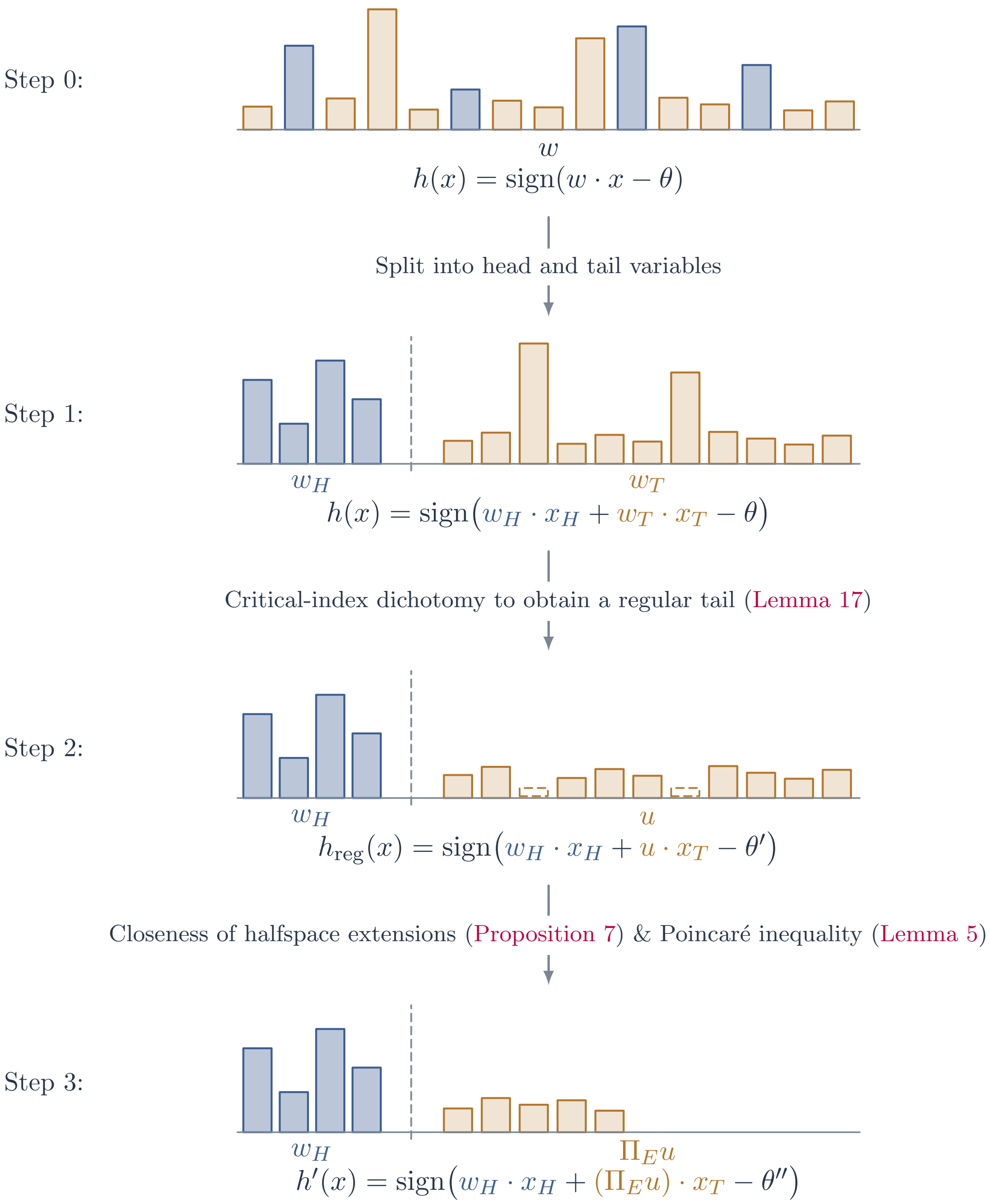

Our construction combines random restrictions with gradient averaging.
Fix $\rho\in(0,1)$, and let $\mu_{\rho,\ast}^{\otimes n}$ be the distribution over restrictions in $\{\pm1,\ast\}^n$ that independently leaves each coordinate free with probability $1-\rho$ and otherwise fixes it to a uniformly random sign.
For a restriction $\bz\sim\mu_{\rho,\ast}^{\otimes n}$, let $\bX_{\rho,f}\in\R^n$ be the average discrete gradient of $f_{\bz}$, padded with zeros on the fixed coordinates.
Equivalently, $\bX_{\rho,f}$ is the degree-$1$ Fourier vector of $f_{\bz}$.
We define the \emph{Boolean influence PCA matrix} by
\[
    M_\rho(f)
    :=
    \Ex\bigl[\bX_{\rho,f}\,\bX_{\rho,f}^\top\bigr]\,.
\]
See~\Cref{fig:query-construction} for an illustration of this construction. 
This construction has two useful properties:
\begin{itemize}
    \item Random restrictions allow higher-degree terms to contribute through linear coefficients.
    For example, fixing $x_2=a$ and $x_3=b$ in $x_1x_2x_3$ leaves $ab\,x_1$.
    Its linear coefficient averages to zero over the random signs $a,b$, but its square is always one, so its contribution survives in the outer product.

    \item Averaging the gradient within each restricted subcube controls the trace.
    Indeed, the resulting vector consists of the degree-$1$ Fourier coefficients of a bounded function, so Parseval's identity gives $\|\bX_{\rho,f}\|_2^2\le1$.
    Consequently, $\Tr(M_\rho(f))\le1$, independently of $n$.
\end{itemize}
We now explain how this matrix determines a common search space and how every halfspace can be transformed into one whose normal lies in that space; see~\Cref{fig:candidate-halfspace-overview} for an illustration. 

\paragraph{Step 1: The Head and Tail Variables.}

For the analysis, we work with the smoothed target $f^{\mathrm{sm}}:=\T_{\sqrt\rho}f$.
Peres' noise-stability bound for halfspaces (\Cref{prop:peres}) ensures that this changes every halfspace correlation only slightly when $\rho$ is close to one.

A short calculation shows that the diagonal entries of $M_\rho(f)$ are proportional to the coordinate influences of $f^{\mathrm{sm}}$.
Thresholding these entries partitions the coordinates into a head $H$ of high-influence variables and a tail $T$ of low-influence variables.
We then let $E$ be the span of the eigenvectors of the tail principal submatrix $M_\rho(f)[T,T]$ whose eigenvalues exceed a suitable threshold.
The trace bound ensures that both $|H|$ and $\dim(E)$ are bounded independently of $n$. 
Note that both $H$ and $E$ depend only on the target $f$.
For an arbitrary candidate halfspace $h(x)=\sign(w\cdot x-\theta)$, we will preserve its head weights $w_H$ and replace its tail weights $w_T$ by a vector in $E$, with little loss in correlation. 

\paragraph{Step 2: Regularizing the Tail.}

We first replace $w_T$ by a \emph{regular} vector $u$, meaning that no individual coordinate is large relative to its Euclidean norm. (See \Cref{def:regular} for the formal definition.) 
A standard critical-index dichotomy~\cite{Servedio07,OST22} gives two cases:
\begin{itemize}
    \item Removing a small set of the largest tail weights leaves a regular vector.
    Restricting the corresponding coordinates then gives a halfspace with a regular tail.

    \item Keeping only a small set of the largest tail weights, together with the head weights, gives a good approximation to $h$.
    Restricting these tail coordinates then gives a halfspace depending only on the head.
\end{itemize} 
In either case, the restricted coordinates have low influence on $f^{\mathrm{sm}}$.
We can therefore choose one restriction that loses little correlation, averaged over all head assignments.
This preserves the original head weights and changes only the tail and the threshold. 
We thus obtain a halfspace 
\[
    h_{\mathrm{reg}}(x)
    =
    \sign(w_H\cdot x_H+u\cdot x_T-\theta')
\]
where $u$ is zero or sufficiently regular. 
Looking ahead, regularity will allow us to replace Boolean tail inputs by Gaussian inputs while approximately preserving correlation with the smoothed target $f^{\mathrm{sm}}$. 

\paragraph{Step 3: From Boolean to Gaussian, and Back.}

We want to show that projecting $u$ onto $E$, the leading eigenspace of the tail matrix $M_\rho(f)[T,T]$ from Step~1, loses little correlation.
This is the intuition behind the Gaussian algorithms~\cite{DMN21,DKKTZ21,DKKTZ24}, and regularity gives us reason to expect it here as well: since the tail is regular, the linear form $u\cdot\bx_T$ is approximately Gaussian.
But this observation alone does not let us replace Boolean tail inputs by Gaussian ones while preserving correlation with an arbitrary target $f^{\mathrm{sm}}$.

\begin{figure}[t]
    \centering
    \resizebox{\textwidth}{!}{
\begingroup
\definecolor{exInk}{HTML}{26364A}
\definecolor{exMuted}{HTML}{637181}
\definecolor{exLine}{HTML}{A3A3A3}
\definecolor{exBlue}{HTML}{3C5E91}
\definecolor{exOchre}{HTML}{B4772C}
\definecolor{exTeal}{HTML}{287F79}

\tikzset{
  ex space/.style={x={(.573576cm,-.384569cm)},
    y={(.819152cm,.269278cm)},z={(0cm,.882948cm)}},
  ex edge/.style={draw=exMuted!70,line width=.55pt},
  ex back/.style={draw=exMuted!38,densely dotted,line width=.5pt},
  ex arrow/.style={-latex,draw=exLine,line width=.7pt},
  ex title/.style={font=\normalfont\normalsize,text=black,inner sep=0pt},
  ex formula/.style={font=\normalfont\normalsize,text=black,inner sep=0pt},
}

\pgfmathdeclarefunction{ext}{3}{%
  \pgfmathparse{2*sqrt(3+(#1*#1+#2*#2)/2)*
    sin(asin(3*sqrt(3)*((#1*#1*#2-#2*#2*#2/3)/sqrt(6)+2*#3)/
      (2*(3+(#1*#1+#2*#2)/2)^1.5))/3)}%
}
\pgfmathdeclarefunction{exx}{3}{%
  \pgfmathparse{#1/sqrt(2)+#2/sqrt(6)+ext(#1,#2,#3)/sqrt(3)}%
}
\pgfmathdeclarefunction{exy}{3}{%
  \pgfmathparse{-#1/sqrt(2)+#2/sqrt(6)+ext(#1,#2,#3)/sqrt(3)}%
}
\pgfmathdeclarefunction{exz}{3}{%
  \pgfmathparse{-2*#2/sqrt(6)+ext(#1,#2,#3)/sqrt(3)}%
}

\newcommand{\exVertices}[1]{%
  \foreach \x in {-1,1}{\foreach \y in {-1,1}{\foreach \z in {-1,1}{%
    \pgfmathtruncatemacro{\positive}{\x+\y+\z>0}
    \ifnum\positive=1\def\exColor{exBlue}\def\exSign{+}%
    \else\def\exColor{exOchre}\def\exSign{-}\fi
    \node[circle,fill=\exColor,draw=white,line width=.3pt,
      inner sep=0pt,minimum size=3mm,text=white,
      font=\scriptsize]
      at (\x,\y,\z) {$\exSign$};
  }}}%
}
\newcommand{\exCube}[1]{%
  \draw[ex back] (-1,1,-1)--(1,1,-1)
    (-1,1,-1)--(-1,-1,-1) (-1,1,-1)--(-1,1,1);
  \draw[ex edge]
    (-1,-1,-1)--(1,-1,-1)--(1,1,-1)--(1,1,1)--(-1,1,1)
      --(-1,-1,1)--(-1,-1,-1)
    (-1,-1,1)--(1,-1,1)--(1,1,1)
    (1,-1,1)--(1,-1,-1);
  \exVertices{#1}%
}

\newcommand{\exSurface}[2]{%
  \pgfmathtruncatemacro{\central}{abs(#1)<.001}
  \ifnum\central=1
    \tikzset{ex surface/.style={fill=#2,fill opacity=.21,
      draw=#2!90,line width=.75pt}}
  \else
    \tikzset{ex surface/.style={fill=#2,fill opacity=.05,
      draw=#2!50,line width=.4pt}}
  \fi
  \path[ex surface]
    plot[domain=0:360,samples=101,variable=\a]
      ({exx(1.5*cos(\a),2.15*sin(\a),#1)},
       {exy(1.5*cos(\a),2.15*sin(\a),#1)},
       {exz(1.5*cos(\a),2.15*sin(\a),#1)}) -- cycle;
}

\begin{tikzpicture}[font=\normalfont,text=black,line cap=round,line join=round]
\path[use as bounding box] (.3,-.95) rectangle (18.0,13.00);

\begin{scope}[shift={(2.65,6.80)},ex space,scale=1.03]
  \exCube{1}
\end{scope}
\node[ex title] at (2.65,4.63) {Boolean halfspace};
\node[ex formula] at (2.65,3.95)
  {$h(x)=\operatorname{sign}(x_1+x_2+x_3)$};

\draw[ex arrow] (4.45,7.25)
  .. controls (6.15,8.575) and (7.80,10.45) .. (9.70,10.45);
\draw[ex arrow] (4.45,6.35)
  .. controls (6.15,5.025) and (7.80,3.15) .. (9.70,3.15);

\begin{scope}[yshift=.35cm]
\begin{scope}[shift={(13.10,10.10)},ex space,scale=1.03]
  \path[fill=exTeal!10,draw=exTeal!65,line width=.65pt]
    ({-1.8/sqrt(2)-1.8/sqrt(6)},{1.8/sqrt(2)-1.8/sqrt(6)},{3.6/sqrt(6)})
    --({1.8/sqrt(2)-1.8/sqrt(6)},{-1.8/sqrt(2)-1.8/sqrt(6)},{3.6/sqrt(6)})
    --({1.8/sqrt(2)+1.8/sqrt(6)},{-1.8/sqrt(2)+1.8/sqrt(6)},{-3.6/sqrt(6)})
    --({-1.8/sqrt(2)+1.8/sqrt(6)},{1.8/sqrt(2)+1.8/sqrt(6)},{-3.6/sqrt(6)})--cycle;
  \foreach \t in {-.9,0,.9}{%
    \draw[exTeal!20,line width=.3pt]
      ({\t/sqrt(2)-1.8/sqrt(6)},{-\t/sqrt(2)-1.8/sqrt(6)},{3.6/sqrt(6)})
      --({\t/sqrt(2)+1.8/sqrt(6)},{-\t/sqrt(2)+1.8/sqrt(6)},{-3.6/sqrt(6)});
    \draw[exTeal!20,line width=.3pt]
      ({-1.8/sqrt(2)+\t/sqrt(6)},{1.8/sqrt(2)+\t/sqrt(6)},{-2*\t/sqrt(6)})
      --({1.8/sqrt(2)+\t/sqrt(6)},{-1.8/sqrt(2)+\t/sqrt(6)},{-2*\t/sqrt(6)});
  }
  \draw[ex arrow,draw=exBlue,line width=.85pt]
    (.09,.09,.09)--(1.45,1.45,1.45) node[above right,text=black,inner sep=1pt] {$+1$};
  \draw[ex arrow,draw=exOchre,line width=.85pt]
    (-.09,-.09,-.09)--(-1.45,-1.45,-1.45) node[below left,text=black,inner sep=1pt] {$-1$};
  \exCube{0}
\end{scope}
\node[font=\normalfont\small,text=black] at (16.18,7.96) {$z_1+z_2+z_3=0$};
\node[ex title] at (13.10,7.25) {Geometric extension};
\node[ex formula] at (13.10,6.57)
  {$h_{\mathrm G}(z)=\operatorname{sign}(z_1+z_2+z_3)$};
\end{scope}

\begin{scope}[yshift=-.35cm]
\begin{scope}[shift={(13.10,3.50)},ex space,scale=1.03]
  \exSurface{-.5}{exOchre}
  \exSurface{0}{exTeal}
  \exSurface{.5}{exBlue}
  \tikzset{ex edge/.style={draw=exMuted!42,line width=.45pt},
    ex back/.style={draw=exMuted!22,densely dotted,line width=.4pt}}
  \exCube{0}
\end{scope}
\draw[-latex,exOchre!45,line width=.4pt] (11.23,3.915)--(11.643,3.915);
\node[font=\normalfont\small,text=black,anchor=east,inner sep=1pt] at (11.16,3.915)
  {$\widetilde h=-\tfrac12$};
\draw[-latex,exBlue!45,line width=.4pt] (15.34,3.39)--(14.557,3.085);
\node[font=\normalfont\small,text=black,anchor=west,inner sep=1pt] at (15.41,3.39)
  {$\widetilde h=+\tfrac12$};
\draw[-latex,exTeal!85,line width=.55pt] (15.17,1.59)--(14.55,1.30)--(13.946,1.506);
\node[font=\normalfont\small,text=black,anchor=west,inner sep=1pt] at (15.24,1.59)
  {$\widetilde h=0$};
\node[ex title] at (13.10,.79) {Multilinear extension};
\node[ex formula] at (13.10,.01)
  {$\displaystyle\widetilde h(z)=\frac{1}{2}(z_1+z_2+z_3-z_1z_2z_3)$};
\end{scope}
\end{tikzpicture}
\endgroup}
    \caption{%
    Two extensions of the majority halfspace $h(x)=\sign(x_1+x_2+x_3)$.
    The geometric extension is Boolean-valued, while the multilinear extension is real-valued; three level sets of the latter are shown, with the zero level set emphasized.
    }
    \label{fig:halfspace-extensions}
\end{figure}
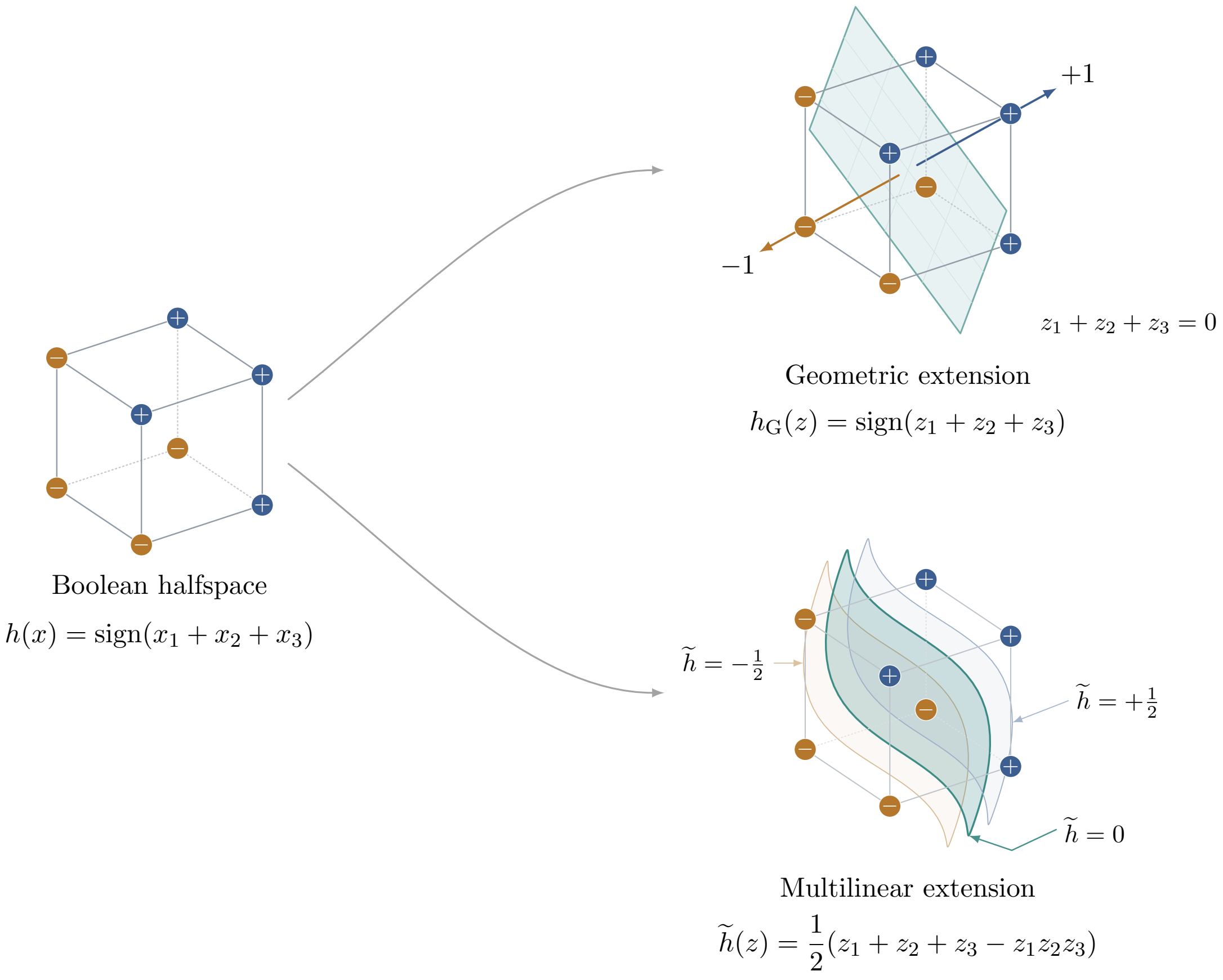

There is a way to pass to Gaussian space that preserves correlation exactly: extend both functions \emph{multilinearly}, by evaluating their Fourier polynomials on real inputs.
For a fixed Boolean head assignment $b$, let $F_b$ and $\widetilde h_b$ be the multilinear extensions of $f^{\mathrm{sm}}(b,\cdot)$ and $h_{\mathrm{reg}}(b,\cdot)$ in the tail coordinates.
Since square-free monomials are orthonormal under both Boolean and Gaussian measures,
\[
    \Ex_{\bx_T\sim\bits^T}[f^{\mathrm{sm}}(b,\bx_T)\,h_{\mathrm{reg}}(b,\bx_T)]
    =\Ex_{\bz\sim N(0,I_T)}[F_b(\bz)\,\widetilde h_b(\bz)]\,.
\]
The difficulty is that $\widetilde h_b$ is a polynomial, rather than a halfspace: averaging out directions from this polynomial does not directly give a halfspace with the projected normal.
There is, however, another natural extension of $h_{\mathrm{reg}}$ to Gaussian space: its \emph{geometric extension}, which simply evaluates the same linear threshold on real inputs,
\[
    h_{b,\G}(z):=\sign(w_H\cdot b+u\cdot z-\theta')\,.
\]
For this extension, the Gaussian averaging argument gives precisely the projection we want.
The two extensions agree on the Boolean cube but generally differ on Gaussian inputs; see \Cref{fig:halfspace-extensions}.
This is where regularity of the tail from the previous step is crucial: it ensures that they are close in Gaussian $L^2$ norm. 
Mossel and Neeman~\cite[Lemma~7.3]{MN15} established such a result with inverse-logarithmic dependence on the regularity parameter. 
\Cref{thm:boolean-gaussian-bridge} gives an exponential quantitative improvement: for a $\tau$-regular tail, we obtain the polynomial bound 
\[
    \vabs{\widetilde h_b-h_{b,\G}}_{L^2(N(0,I_T))} =O(\tau^{1/4})\,.
\]
We also show that the exponent $1/4$ is optimal, with a matching lower bound for the majority function (\Cref{prop:boolean-gaussian-bridge-sharpness}).

Returning to the main argument, we first extend both functions multilinearly in the tail coordinates, preserving their correlation exactly.
We then replace only the halfspace's multilinear extension by its geometric extension. 
Since $\|F_b\|_2\le1$, Cauchy--Schwarz and the comparison above bound the change in correlation by $O(\tau^{1/4})$.
This bound holds for every head assignment, so it also holds after averaging over the Boolean head inputs.

Next, we carry out the Gaussian projection argument described at the start of Step~3.
Write $u=u_E+u_\perp$, where $u_E:=\Pi_Eu$.
The Gaussian halfspace depends on the discarded directions only through the scalar $u_\perp\cdot\bz$, which is independent of $\Pi_E\bz$, where $\bz\sim N(0,I_T)$.
The one-dimensional Gaussian Poincar\'e inequality together with additional properties of the subspace $E$ let us average out the direction of $u_\perp$ with little loss in correlation.
The averaged classifier is a mixture of halfspaces with tail normal $u_E$ and different thresholds.
Choosing one threshold after averaging over all head assignments gives a single halfspace with the original head weights and at least the mixture's correlation. 

Finally, we return to the Boolean cube.
This requires $u_E$ to be regular, which does not follow from the regularity of $u$ alone.
Here the small diagonal entries of the tail matrix $M_\rho(f)[T,T]$ are important: choosing the diagonal cutoff sufficiently small relative to the eigenvalue cutoff ensures that every vector in $E$ is regular.
We can therefore apply the comparison between extensions once more and obtain 
\[
    h'(x)
    =
    \sign(w_H\cdot x_H+(\Pi_Eu)\cdot x_T-\theta'')
\]
with little additional loss in correlation.

\paragraph{Putting Everything Together.}

We can choose parameters appropriately to make the total correlation loss sufficiently small and simultaneously have $D:=|H|+\dim(E)=O(\eps^{-14})$.
The learner therefore searches over halfspaces with arbitrary head weights and tail weights in $E$.
This is a search in only $D$ dimensions, and the argument above guarantees that this restricted class contains a nearly best-fitting halfspace.

The algorithm finds $H$ and $E$ by estimating $M_\rho(f)$ using membership queries.
(Note that membership queries beyond random examples are used only to estimate the matrix.)
The estimation error is easily handled via standard concentration bounds, so suitable thresholds applied to the estimated matrix give the required head and tail variables as well as the leading eigenspace $E$.  
The learner then performs empirical risk minimization using $m=O(D/\eps^2)$ fresh uniformly random labeled examples. 
The search takes $m^{O(D)}=2^{O(\eps^{-14}\log(1/\eps))}$ time. 
Matrix estimation and feature construction take time polynomial in $n$ and $1/\eps$, followed by a search whose cost depends only on $\eps$.
Adding these costs gives total running time $\poly(n,1/\eps)+2^{O(\eps^{-14}\log(1/\eps))}$, with total query complexity $O(n\eps^{-28})$.

\subsection{Related Work}
\label{subsec:related-work}

We give a brief overview of related work, starting with previous work on agnostically learning halfspaces under the Gaussian distribution. 

\paragraph{Proper Agnostic Learning of Halfspaces over Gaussian Space.}
Analogues of both our main results were previously known under the standard Gaussian distribution $N(0,I_n)$.
Over this marginal distribution, polynomial regression already achieved error $\OPT+\eps$ with an improper hypothesis~\cite{KKMS:08} using $n^{\poly(1/\eps)}$ samples. 
Subsequent work obtained proper learners in both access models:

\begin{itemize}
    \item 
    Diakonikolas et al.~\cite{DKKTZ24} gave efficient query-based agnostic learners for the broad class of \emph{multi-index models} under Gaussian marginals.
    As a consequence, they obtain a proper halfspace learner with error $\OPT+\eps$, using $\poly(n,1/\eps)$ samples and queries and running in time $\poly(n,1/\eps)+2^{\poly(1/\eps)}$.
    Our \Cref{thm:main} obtains an analogous guarantee over the Boolean hypercube. 
    
    \item 
    Diakonikolas, Kane, Kontonis, Tzamos, and Zarifis~\cite{DKKTZ21} gave a proper agnostic learner with error $\OPT+\eps$ and running time $n^{\poly(1/\eps)}$.
    (See Tikhonov and Vasilyan~\cite{TV26} for a recent improvement to the running time and extension to proper learning of functions of a bounded number of halfspaces.) 
    Our \Cref{thm:proper-sample} gives a Boolean analogue of the proper learning guarantee of~\cite{DKKTZ21}. 
\end{itemize}

Both Gaussian learners use spectral information about influential directions to reduce the search for a nearly optimal halfspace to a low-dimensional subspace.
In~\cite{DKKTZ21}, these directions are extracted from a low-degree polynomial fitted to the labels; in~\cite{DKKTZ24}, they are extracted from a smoothed version of the target.
In both cases, the relevant matrix is the second moment of the gradient, and its top eigenspace contains the normal of a nearly optimal halfspace. 
A similar gradient-based PCA approach appears in the work of De, Mossel, and Neeman~\cite{DMN21} on robust testing of \emph{linear $k$-juntas} over Gaussian space, a class that includes halfspaces when $k=1$.

Our query-based algorithm underlying~\Cref{thm:main} gives a Boolean counterpart of this approach.
The Boolean setting presents additional challenges absent in the Gaussian setting: for example, arbitrary rotations do not preserve the uniform distribution over $\bn$, and powerful technical tools such as \emph{anti-concentration} and the \emph{localization technique}~\cite{ABL17,DKS18} are no longer available. 
(See~\cite{chandrasekaran2026fully} for a broader discussion of the obstacles to extending continuous learning algorithms to discrete distributions.)
Indeed, obtaining a similar dimension reduction over $\bn$ requires several additional ingredients, including a discrete influence matrix, critical-index analysis, and a quantitative comparison between the geometric and multilinear Gaussian extensions of regular halfspaces. 

\paragraph{Robust Learning of Halfspaces.}

A related line of work gives efficient learning algorithms with relaxed agnostic guarantees of the form $g(\OPT)+\eps$, where $g(t)\to0$ as $t \to 0$. 
Under the uniform distribution over $\bn$, O'Donnell and Servedio~\cite{OS11:chow} obtain error $O(\OPT^c)+\eps$ for an absolute constant $c>0$.
De, Diakonikolas, Feldman, and Servedio~\cite{DDFS14} improved the running-time dependence on $\eps$, at the cost of a weaker dependence on $\OPT$. 
Both algorithms use uniformly random examples and run in time polynomial in $n$ for every fixed $\eps$, but super-polynomial in $1/\eps$.
Recently, Chandrasekaran, Klivans, Stavropoulos, and Vasilyan~\cite{chandrasekaran2026fully} gave a proper learner with error $O(\OPT^{c'})+\eps$ for an absolute constant $c'>0$ and running time $\poly(n,1/\eps)$.
Similar relaxed agnostic guarantees have also been studied extensively under Gaussian and other continuous marginal distributions; see the related-work discussion in~\cite{chandrasekaran2026fully} for an overview. 

\paragraph{Learning with Queries.}

The power of membership queries in agnostic learning differs substantially between the distribution-free and distribution-specific settings. 
For concept classes on the Boolean hypercube, Feldman~\cite{Feldman09MQ} showed that distribution-free agnostic learners using membership queries can be simulated using random examples alone, with polynomial overhead. 
He also exhibited a computational advantage for membership queries under the uniform Boolean distribution, assuming the existence of one-way functions, and asked whether such an advantage could be established for natural classes such as halfspaces. 
Gopalan, Kalai, and Klivans~\cite{GKK08open} asked the stronger quantitative question of whether halfspaces admit an agnostic query learner running in time $\poly(n,1/\eps)$. 
Diakonikolas et al.~\cite{DKKTZ24} obtained improved agnostic learning guarantees using membership queries in Gaussian space; our \Cref{thm:main} improves the dimension dependence over polynomial regression in the Boolean setting.
Our result also makes progress on the stronger question posed in~\cite{GKK08open}: it achieves polynomial query complexity and a running-time exponent in $n$ independent of $\eps$, while the fully polynomial running time remains open.  

Membership queries also underlie several classical learning algorithms under the uniform Boolean distribution, including the Goldreich--Levin algorithm for parities~\cite{GoldreichLevin:89} and the Kushilevitz--Mansour algorithm for decision trees~\cite{KushilevitzMansour:93}. 
Building on these techniques, Gopalan, Kalai, and Klivans~\cite{GKK:08} gave an efficient improper agnostic learner for decision trees under the uniform Boolean distribution. 
In the distribution-free setting, recent works by  Alman, Nadimpalli, Patel, and
Servedio~\cite{ANPS25walks,ANPS25exact} also exploit the power of membership and equivalence queries to design new algorithms for learning DNF formulas. 

\paragraph{Proper Learning of Other Classes.}

For monotone Boolean functions under the uniform distribution, the Fourier concentration theorem of Bshouty and Tamon~\cite{BshoutyTamon:96}, combined with polynomial regression~\cite{KKMS:08}, yields an improper agnostic learner using random examples and running in time \smash{$2^{\wt{O}(\sqrt n)}$} for every fixed $\eps>0$. 
Lange, Rubinfeld, and Vasilyan~\cite{LRV22} used local correction to obtain a \emph{proper} learner with error $3\OPT+\eps$ and running time \smash{$2^{\wt{O}(\sqrt n)}$}. 
Lange and Vasilyan~\cite{LV23} subsequently achieved the fully agnostic guarantee $\OPT+\eps$ with the same asymptotic running time. 
Proper agnostic learners are also known for Lipschitz convex functions~\cite{FPMMN26} and for restricted families of convex bodies~\cite{DNS24,de2024sparsifying,nadimpalli2026optimal} under the Gaussian measure.

\subsection{Acknowledgements and AI Methodology}
\label{subsec:acks}

A.D.~is supported by~NSF~CCF~2045128 and the UK AI Security Institute (AISI). 
E.M.~is supported by NSF grant DMS-2031883, by the Vannevar Bush Faculty Fellowship ONR-N00014-
20-1-2826, and by a Simons Investigator Award in Mathematics.
S.N.~is partially supported by E.M.'s Vannevar Bush Faculty Fellowship ONR-N00014-20-1-2826. 
S.N.~thanks Ronitt Rubinfeld for asking whether \Cref{thm:main} could yield a sample-based agnostic learning algorithm, a question that led to \Cref{thm:proper-sample}. 

The authors constructed the discrete matrix $M_\rho(f)$ and designed a proper agnostic learning algorithm for halfspaces without assistance from  large language models (LLMs). 
The algorithm ran in time $O_\eps(1)\cdot\poly(n)$, but its dependence on $\eps^{-1}$ was doubly exponential.
The quantitative bottleneck was the Boolean--Gaussian comparison of Mossel and Neeman~\cite[Lemma~7.3]{MN15}.
We then asked an LLM whether this inequality could be sharpened, and it supplied the proof of \Cref{thm:boolean-gaussian-bridge} presented in \Cref{app:robust-chow}. 
This improved our algorithm's running-time dependence on $\eps^{-1}$ from doubly exponential to singly exponential, giving the bound in \Cref{thm:main}. 

We also used LLMs for copyediting, proofreading, generating diagrams, and tracking the powers of $n$ and $\eps^{-1}$ in our complexity bounds. 
The authors take full responsibility for the correctness and contents of this paper.

\section{Preliminaries}
\label{sec:preliminaries}

We write $\|\cdot\|_2$ for the Euclidean $\ell_2$-norm, and denote the unit sphere in $\R^n$ by $\mathbb{S}^{n-1}$. 
We write $[n]:=\{1,\dots,n\}$.
For $J\subseteq[n]$, write $J^c:=[n]\setminus J$ for the complement of $J$.
Throughout, letters such as $c$ and $C$ denote positive universal constants whose values may change from appearance to appearance unless they are explicitly fixed.
In complexity bounds, $\log n$ denotes $\max\{1,\log n\}$; the base of the logarithm is immaterial. 

\paragraph{Linear Algebra.} 
For $S\subseteq[n]$ and $x\in\R^n$, write $x_S:=(x_i)_{i\in S}$.
For $S,T\subseteq[n]$ and $M\in\R^{n\times n}$, write $M[S,T]$ for the submatrix of $M$ with rows indexed by $S$ and columns indexed by $T$.
In particular, $M[S,S]$ is the principal submatrix of $M$ indexed by $S$.
For a linear subspace $E\subseteq\R^n$, let $\Pi_E$ denote the orthogonal projector onto $E$.

\paragraph{Probability.}
We use boldfaced letters, such as $\bx$ and $\bX$, to denote random variables, which may be real-, vector-, or matrix-valued.
The intended type will be clear from context.
Given a distribution $\calD$, we write $\bx\sim\calD$ to indicate that $\bx$ is distributed according to $\calD$.
Given a probability space $(\Omega,\mu)$ and $p\in[1,\infty)$, we write $L^p(\Omega,\mu)$ for the space of measurable functions $f:\Omega\to\R$ satisfying
\[
	\|f\|_{L^p(\Omega,\mu)}
	:=
	\Ex_{\bx\sim\mu}\sbra{|f(\bx)|^p}^{1/p}
	<\infty\,.
\]
When the underlying probability space is clear from context, we abbreviate $L^p(\Omega,\mu)$ and $\|\cdot\|_{L^p(\Omega,\mu)}$ by $L^p$ and $\|\cdot\|_{L^p}$, respectively.
The space $L^2(\Omega,\mu)$ is equipped with the inner product
\[
	\langle f,g\rangle_{L^2(\Omega,\mu)}
	:=
	\Ex_{\bx\sim\mu}[f(\bx)\,g(\bx)]\,,
	\qquad\text{so that}\qquad
	\|f\|_{L^2(\Omega,\mu)}^2
	=
	\langle f,f\rangle_{L^2(\Omega,\mu)}\,.
\]
We often suppress the measure and write $\langle f,g\rangle$ when the ambient space is understood.
For bounded functions $f$ and $g$ on the same probability space, define their correlation by
\[
	\corr_f(g):=\Ex[f\,g]\,.
\]
We will almost always use correlation for Boolean functions on $\bits^n$ under the uniform measure.
For such functions $f$ and $g$, correlation and distance satisfy
\[
	\corr_f(g)
	=
	1-2\err_f(g)\,,
	\qquad\text{or equivalently}\qquad
	\err_f(g)
	=
	\frac{1-\corr_f(g)}{2}\,.
\]
We will primarily work with two probability spaces.
On the Boolean hypercube $\bn$, we always use the uniform probability measure and write $L^p(\bn)$ for the corresponding $L^p$ space.
On $\R^n$, we use the standard Gaussian measure $N(0,I_n)$ and write $L^p(N(0,I_n))$ for the corresponding space.

\subsection{Fourier Analysis on the Hypercube}
\label{subsec:fourier}

Our notation and terminology follow \cite{OD14}. 
Recall that the collection of parity functions $\{\chi_S(x) := \prod_{i \in S} x_i \}_{S\sse[n]}$ forms an orthonormal basis for $L^2(\bn)$ with the convention that $\chi_{\emptyset} \equiv 1$. 
As a result, every $f:\bn\to\R$ has the Fourier expansion
\[
	f(x) = \sum_{S \sse [n]} \wh{f}(S) \chi_S(x)\,,
	\qquad\text{where}\qquad 
	\wh{f}(S) := \Ex_{\bx\sim\bits^n}\sbra{f(\bx)\,\chi_S(\bx)}\,.
\]
We will abuse notation and write $\wh{f}(i)=\wh{f}(\{i\})$ for singleton Fourier coefficients.
Recall Parseval and Plancherel's identities, which state that for $f, g: \bn \to \R$, 
\[
	\Ex_{\bx\sim\bits^n}\sbra{f(\bx)^2} = \sum_{S\subseteq[n]}\wh f(S)^2\,,
	\qquad 
	\Ex_{\bx\sim\bn}\sbra{f(\bx)\,g(\bx)} = \sum_{S\sse[n]} \wh{f}(S)\,\wh{g}(S)\,.
\]

Let $\rho \in [0,1]$.
Recall that the \emph{influence} and the \emph{$\rho$-noisy influence} of a coordinate $i \in [n]$ on a function $f:\bn\to\R$ on the Boolean cube are respectively 
\[
	\Inf_i[f] := \sum_{S\ni i} \wh{f}(S)^2
	\qquad\text{and}\qquad  
	\Inf_i^{(\rho)}[f] := \sum_{S\ni i}\rho^{|S|-1}\wh f(S)^2\,,
\]
with the convention $\Inf_i^{(0)}[f]=\wh f(i)^2$. 
Thus the definition at $\rho=0$ retains exactly the degree-$1$ Fourier contribution.

Let $x \in \bn$. 
We define the distribution $N_\rho(x)$ over $\bn$ where each coordinate of $\by \sim N_\rho(x)$ is obtained as follows: 
\[
	\by_i = 
	\begin{cases}
		x_i & \text{with probability}~\rho \\ 
		+1 & \text{with probability}~(1-\rho)/2\\
		-1 & \text{with probability}~(1-\rho)/2
	\end{cases}\,.
\]
The Boolean noise operator $\T_\rho$ is defined by $\T_\rho f(x) = \Ex_{\by\sim N_\rho(x)}\sbra{f(\by)}$. 
It is readily verified that \smash{$\T_\rho f = \sum_{S} \rho^{|S|} \wh{f}(S) \chi_S$}. 
Finally, for $J\subseteq[n]$, $z\in\bits^J$, and $y\in\bits^{J^c}$, let $x^{z,y}\in\bits^n$ be defined coordinatewise by
\[
	(x^{z,y})_i
	=
	\begin{cases}
		z_i, & i\in J\,,\\
		y_i, & i\in J^c\,.
	\end{cases}
\]
The restriction $f_z:\bits^{J^c}\to\R$ is defined by $f_z(y):=f(x^{z,y})$.

We require the following observation: resampling a set of low-influence coordinates changes correlation with a bounded function by only a small amount.

\begin{lemma}[Coordinate resampling]
	\label{lem:coordinate-resampling}
	Let $f:\bits^n\to[-1,1]$, let $J\subseteq[n]$, and let $g:\bits^n\to[-1,1]$.
	Define
	\[
		g_J(x)
		:=
		\Ex_{\by_J\sim\bits^J}\sbra{g(\by_J,x_{-J})}\,.
	\]
	Then
	\[
		\abs{
		\Ex_{\bx}\sbra{f(\bx)\,g(\bx)}
		-
		\Ex_{\bx}\sbra{f(\bx)\,g_J(\bx)}
		}
		\le
		\sqrt{\sum_{i\in J}\Inf_i[f]}\,.
	\]
\end{lemma}

\begin{proof}
	The map $u\mapsto u_J$ is self-adjoint.
	Hence, by Cauchy--Schwarz,
	\[
		\abs{\Ex\sbra{f\,g}-\Ex\sbra{f\,g_J}}
		=
		\abs{\Ex\sbra{(f-f_J)\,g}}
		\le
		\vabs{f-f_J}_{L^2}\,.
	\]
	Moreover,
	\[
		\vabs{f-f_J}_{L^2}^2
		=
		\sum_{S:\,S\cap J\ne\emptyset}\wh f(S)^2
		\le
		\sum_{i\in J}\Inf_i[f]\,,
	\]
	completing the proof. 
\end{proof}

\subsection{Gaussian Space and Multilinear Extensions}
\label{subsec:gaussian}

Given a function $f:\bn\to\R$, its multilinear extension $\wtf:\R^n\to\R$ is given by
\[
	\wt f(x)=\sum_{S\subseteq[n]}\wh f(S)\prod_{i\in S}x_i\,.
\]
The same square-free monomials are orthonormal under $\bz\sim N(0,I_n)$.
Consequently
\[
	\Ex_{\bx\sim\bits^n}\sbra{f(\bx)\,g(\bx)}
	=
	\Ex_{\bz\sim N(0,I_n)}
	\sbra{\wt{f}(\bz)\,\wt{g}(\bz)}
\]
for every pair of functions $f,g:\bits^n\to\R$.

Suppose $f \in L^1(N(0,I_n))$ and $\rho \in [0,1]$.
The Ornstein--Uhlenbeck operator $\U_\rho$ is defined as 
\[
	\U_\rho f(x) := \Ex_{\bz\sim N(0, I_n)}\sbra{f\pbra{\rho x + \sqrt{1-\rho^2}\bz}}\,.
\]
It is readily checked that $\U_\rho$ acts on square-free monomials similarly to the Boolean noise operator $\T_\rho$~\cite[Chapter~11]{OD14}.  
In particular, for $f:\bits^n\to\R$, we have 
\[
	\U_\rho \wtf(x) = \sum_{S\sse[n]}\rho^{|S|}\wh{f}(S)\prod_{i\in S} x_i\,.
\]
Finally, $\U_\rho$ is self-adjoint in $L^2(N(0,I_n))$; in other words, for $f, g \in L^2(N(0,I_n))$, we have 
\[
	\Ex_{\bz\sim N(0,I_n)}\sbra{f(\bz)\,\U_\rho g(\bz)}
	= 
	\Ex_{\bz\sim N(0,I_n)}\sbra{\U_\rho f(\bz)\, g(\bz)}\,.
\]

We will require the following averaging operator used by \cite{DMN21}. 
Let $f \in L^1(N(0,I_n))$. 
For a subspace $E\subseteq\R^n$, define the Gaussian averaging operator
\[
	\calA_E f(x)
	:=
	\Ex_{\bz\sim N(0,\Pi_{E^\perp})}
	\sbra{f(\Pi_Ex+\bz)}\,,
\]
where the Gaussian is supported on $E^\perp$. 
Note that $\calA_E f(x)$ only depends on $\Pi_E x$.
We note that $\calA_E$ is self-adjoint in $L^2(N(0,I_n))$; see \cite[Lemma~2.2]{DMN21} for a proof. 

\subsection{Halfspaces}
\label{subsec:halfspaces-prelims}

Throughout the paper, we use 
\[
	\sign(t)=
	\begin{cases}
		+1 & t\ge 0\\
		-1 & t<0
	\end{cases}\,.
\]
A non-constant halfspace is written as $h(x)=\sign(w\cdot x-\theta)$
with normal vector $w$ and threshold $\theta \in \R$. 
Given a halfspace $h\isafunc$ of the form $h(x)=\sign(w\cdot x-\theta)$, we use the same defining inequality to extend $h$ geometrically to a halfspace on $\R^n$:
\[
	h_{\G}(z):=\sign(w\cdot z-\theta)\,,
	\qquad z\in\R^n\,.
\]
We also identify $h$ with its unique multilinear extension to $\R^n$, namely
the multilinear polynomial $\wt{h}$ (as defined in \Cref{subsec:gaussian}) which agrees with $h$ on $\bn$.

If $E\subseteq\R^n$ is a linear subspace, let $\calH_E$ be the class of affine halfspaces $h:\bits^n\to\bits$ whose normal lies in $E$:
\[
	\calH_E
	:=
	\cbra{x\mapsto\sign(w\cdot x-\theta):w\in E,\ \theta\in\R}\,.
\]
This definition includes both constant halfspaces, obtained by taking $w=0$ and choosing $\theta\le0$ or $\theta>0$.

\subsubsection{Useful Estimates}
\label{subsubsec:halfspace-useful-estimates-prelims}

We use Peres' dimension-free noise-stability bound for halfspaces~\cite{Peres04}; see also \cite[Section~5.5]{OD14}.

\begin{proposition}[Peres' theorem] \label{prop:peres}
	Let $h(x)=\sign(w\cdot x-\theta)$ be a halfspace over $\bn$, and let $\rho\in[0,1]$.
	Then
	\[
		\|\T_\rho h-h\|_{L^1(\bn)}
		\leq 2\sqrt{2(1-\rho)}\,.
	\]
\end{proposition}

We will also require the following consequence of the one-dimensional Gaussian Poincar\'{e} inequality; see, for example, \cite[Section~11.2 and Exercise~11.13]{OD14}.
It bounds the cost of projecting a halfspace using only the derivative in the direction removed from its normal.

\begin{lemma} \label{lem:poincare-projection}
	Let $F:\R^n\to\R$ be smooth with $F$ and $\|\nabla F\|_2$ square-integrable under $N(0,I_n)$.
	Let $E\subseteq\R^n$ be a subspace, and let $h_\G(x)=\sign(w\cdot x-\theta)$.
	All expectations below are over $\bz\sim N(0,I_n)$.
	If $\Pi_{E^\perp}w\ne0$, set $v:=\Pi_{E^\perp}w/\|\Pi_{E^\perp}w\|_2$.
	Then
	\[
		\left|\Ex\sbra{F(\bz)\,h_\G(\bz)}
		-\Ex\sbra{F(\bz)\,\calA_Eh_\G(\bz)}\right|
		\le
		\left(\Ex\sbra{(v\cdot\nabla F(\bz))^2}\right)^{1/2}\,.
	\]
	If $\Pi_{E^\perp}w=0$, the two correlations are equal.
\end{lemma}

\begin{proof}
	The zero-projection case is immediate, so suppose $\Pi_{E^\perp}w\ne0$.
	Let $\mathcal P_v$ average only the Gaussian coordinate along $v$:
	\[
		\mathcal P_vF(y)
		:=\Ex_{\bs\sim N(0,1)}
		\sbra{F\bigl(y-(v\cdot y)v+\bs v\bigr)}\,.
	\]
	This conditional-expectation operator is self-adjoint in Gaussian $L^2$.
	The halfspace $h_\G$ depends on $E^\perp$ only through its coordinate along $v$, so $\mathcal P_vh_\G=\calA_Eh_\G$.
	Consequently, self-adjointness and Cauchy--Schwarz give
	\[
		\left|\Ex\sbra{F(h_\G-\calA_Eh_\G)}\right|
		=\left|\Ex\sbra{(F-\mathcal P_vF)h_\G}\right|
		\le\|F-\mathcal P_vF\|_2\,.
	\]
	Condition on the Gaussian coordinates in $v^\perp$ and apply the one-dimensional Gaussian Poincar\'{e} inequality along $v$.
	Averaging the resulting conditional-variance bound yields
	\[
		\|F-\mathcal P_vF\|_2^2
		\le\Ex\sbra{(v\cdot\nabla F(\bz))^2}\,,
	\]
	which proves the claim.
\end{proof}

\subsubsection{Regular Halfspaces}
\label{subsubsec:regular-prelims}

The following notion of regularity is due to Servedio~\cite{Servedio07}; related notions and decompositions were subsequently used in work on testing and fooling halfspaces~\cite{MORS10,GOWZ10}. 

\begin{definition}[Regularity] \label{def:regular}
	Let $\tau>0$.
	We declare the zero vector to be $\tau$-regular. 
	An arbitrary non-zero vector $w\in\R^n$ is $\tau$-regular if it satisfies
	\[
		\left\|w\right\|_\infty\leq\tau\cdot\|w\|_2\,.
	\]
	A halfspace $h(x)=\sign(w\cdot x-\theta)$ is $\tau$-regular if its normal vector $w$ is $\tau$-regular. 
\end{definition}

Note that a halfspace $h(x)=\sign(w\cdot x-\theta)$ on $\bn$ has two distinct extensions to $\R^n$: the geometric extension $h_{\G}$, obtained from the same defining inequality, and the multilinear extension $\wt{h}$, obtained from the Fourier expansion of $h$.
We will require the following result, which shows that these two extensions are close in $L^2(N(0,I_n))$ when $h$ is regular.
It provides a bridge between the Boolean and Gaussian settings.
We will use it repeatedly throughout the rest of the paper to transfer correlation estimates between them.

\begin{restatable}[Closeness of multilinear and geometric extensions for regular halfspaces]{proposition}{BooleanGaussianBridge}
\label{thm:boolean-gaussian-bridge}
	There is a universal constant $C$ such that the following holds.
	Let
	\[
		h(x)=\sign(w\cdot x-\theta)
	\]
	be a $\tau$-regular halfspace, where $\tau \in (0,1]$.
	Let $h_\G(\bz)=\sign(w\cdot \bz-\theta)$ and let $\wt h$ be the multilinear extension of the Boolean halfspace.
	Then
	\[
		\Ex_{\bz\sim N(0,I_n)}
		\sbra{\pbra{\wt{h}(\bz)-h_\G(\bz)}^2}
		\le
		C\tau^{1/2}\,.
	\]
	More precisely, if $\|w\|_2=1$ and $\beta_w:=(\sum_iw_i^4)^{1/2}$, then
	\[
		\|\wt h-h_\G\|_{L^2(N(0,I_n))}^2\le C\beta_w^{1/2}.
	\]
	Note that if $w=0^n$ the left-hand side is zero; otherwise $w$ and $\theta$ may be normalized by the same positive factor.
\end{restatable}

The proof of \Cref{thm:boolean-gaussian-bridge} is given in \Cref{app:robust-chow}.
Mossel and Neeman~\cite[Lemma~7.3]{MN15} proved a weaker version of this result using the invariance principle, with only logarithmic dependence on the regularity parameter $\tau$. 
\Cref{thm:boolean-gaussian-bridge} thus gives an exponential quantitative improvement over~\cite[Lemma~7.3]{MN15}.
We also note that the exponent above: the majority function gives a matching lower bound; see~\Cref{prop:boolean-gaussian-bridge-sharpness}.

As mentioned above, we will repeatedly use \Cref{thm:boolean-gaussian-bridge} to pass between the Boolean and Gaussian settings when working with regular halfspaces.
To see this, let $h:\bn\to\bits$ be a $\tau$-regular halfspace, let $h_\G$ be its geometric Gaussian extension, and let $f:\bn\to[-1,1]$. 
Because square-free monomials are orthonormal under both the uniform measure on $\bn$ and the standard Gaussian measure, the correlation between $f$ and $h$ is 
\begin{align*}
	\Ex_{\bx\sim\bn}\sbra{f(\bx)\,h(\bx)}
	&= \Ex_{\bz\sim N(0,I_n)}\sbra{\wtf(\bz)\,\wt h(\bz)} \\
	&= \Ex_{\bz\sim N(0,I_n)}\sbra{\wtf(\bz)\,h_\G(\bz)}
	 + \Ex_{\bz\sim N(0,I_n)}\sbra{\wtf(\bz)\,\pbra{\wt h(\bz)-h_\G(\bz)}}\,.
\end{align*}
The second term is small by Cauchy--Schwarz and \Cref{thm:boolean-gaussian-bridge}:
\[
	\left|
	\Ex_{\bz\sim N(0,I_n)}
	\sbra{\wtf(\bz)\,\pbra{\wt h(\bz)-h_\G(\bz)}}
	\right|
	\leq
	\|\wtf\|_{L^2(N(0,I_n))}\,\|\wt h-h_\G\|_{L^2(N(0,I_n))}
	\leq C\tau^{1/4}\,.
\]
Here we used $\|\wtf\|_{L^2(N(0,I_n))}=\|f\|_{L^2(\bn)}\leq1$.
Consequently,
\[
	\left|
	\Ex_{\bx\sim\bn}\sbra{f(\bx)\,h(\bx)}
	-
	\Ex_{\bz\sim N(0,I_n)}\sbra{\wtf(\bz)\,h_\G(\bz)}
	\right|
	\leq C\tau^{1/4}\,.
\]
The same argument also gives the reverse passage: if $h_\G$ has a regular normal and $h$ is the Boolean halfspace defined by the same inequality, then we may replace $h_\G$ by the multilinear extension $\wt h$ and convert the resulting Gaussian inner product back into a Boolean correlation.

\subsubsection{The Critical Index}
\label{subsubsec:critical-indices}

We use the critical-index dichotomy for halfspaces, originating in the work of Servedio~\cite{Servedio07}. The form below follows directly from~\cite[Lemma~A.3]{OST22}.

\begin{definition}[Critical index]
    \label{def:crit-index}
    Let $\tau\in(0,1)$ and order $w\in\R^m$ so that $|w_1|\ge\cdots\ge|w_m|$.
    The $\tau$-critical index of $w$ is the least $k\in\{0,\ldots,m\}$ for which
    \[
        \max_{j>k}|w_j|\le\tau\left(\sum_{j>k}w_j^2\right)^{1/2}.
    \]
    We take the maximum over an empty set to be zero, so the empty tail is regular and the critical index is always defined.
\end{definition}

For $\nu\in(0,1/2)$, write $\Lambda_\nu:=\log(e+\log(1/\nu))$.

\begin{proposition}[Critical-index dichotomy]
    \label{thm:critical-index}
    There is a universal constant $C_{\mathrm{crit}}\ge1$ such that the following holds.
    Let $\tau,\nu\in(0,1/2)$, let $w\in\R^m$ have coordinates ordered by decreasing absolute value, and put
    \[
        K:=\left\lceil C_{\mathrm{crit}}\tau^{-2}\log(1/\nu)\Lambda_\nu\right\rceil.
    \]
    Let $\ell$ be the $\tau$-critical index of $w$.
    If $\ell\le\min\{K,m\}$, the tail after the first $\ell$ coordinates is zero or $\tau$-regular.
    Otherwise $K<m$, and truncation to the first $K$ coordinates satisfies
    \[
        \sup_{\theta\in\R}\Prx_{\bx\sim\bits^m}\left[
        \sign(w\cdot\bx-\theta)
        \ne\sign\left(\sum_{i=1}^K w_i\bx_i-\theta\right)
        \right]\le\nu.
    \]
\end{proposition}

\begin{proof}
    The first case is the definition, so suppose $\ell>K$.
    In the terminology of~\cite[Definition~3.1]{OST22}, $w$ is not $(K,\tau)$-regular: if deleting a set $S$ of at most $K$ coordinates left a regular tail, let $j$ be the first undeleted coordinate. Restoring any deleted coordinates after $j$ can only increase the tail norm without increasing its largest absolute coefficient. Thus the suffix starting at $j$ would be regular, giving $\ell\le j-1\le|S|\le K$, a contradiction.

    At error parameter $c\nu$ for a sufficiently small universal $c>0$, \cite[Lemma~A.3]{OST22} gives an integer $k_0\le K$ when $C_{\mathrm{crit}}$ is sufficiently large. 
    The vector $w$ is also not $(k_0,\tau)$-regular. 
    Apply that lemma with the replacement vector equal to $w$ on the first $K$ coordinates and zero elsewhere. 
    It agrees with $w$ on the first $k_0$ coordinates and has no larger norm on the remaining coordinates. 
    Fully independent uniform signs satisfy the lemma's bounded-independence hypothesis. 
    Its conclusion holds for every threshold with this same replacement vector, with error at most $\nu$ by the choice of $c$.
    To match its indicator convention exactly, apply it to $-w$ and $-\theta$ and use $\sign(w\cdot x-\theta)=2\cdot\mathbf{1}\{-w\cdot x\le-\theta\}-1$. \end{proof}

\subsection{Efficient Search over Halfspaces in Fixed Dimension}
\label{subsec:fixed-dimensional-erm}

We record a standard learning primitive for later use. 
Let $\Phi:\bits^n\to\R^D$ be any map. 
The following proposition gives an efficient way to search over all classifiers of the form $h\circ\Phi$, where $h$ is an affine halfspace on $\R^D$. 

\begin{proposition}[Efficient search over affine halfspaces]
	\label{prop:fixed-dimensional-erm}
	Let $S=((x_1,y_1),\ldots,(x_m,y_m))$ be a labeled sample from $\bits^n\times\bits$, and suppose that the mapped points $\Phi(x_1),\ldots,\Phi(x_m)$ are given.
	Define
	\[
		\widehat{\err}_S(h\circ\Phi)
		:=
		\frac{1}{m}
		\sum_{j=1}^m
		\mathbf{1}\{h(\Phi(x_j))\ne y_j\}\,.
	\]
	In time $m^{O(D)}$, one can find an affine halfspace $h_S$ on $\R^D$ minimizing $\widehat{\err}_S(h\circ\Phi)$ over all affine halfspaces $h$ on $\R^D$.
	Moreover, let $P$ be any distribution on $\bits^n\times\bits$, and let $\bS=((\bx_1,\by_1),\ldots,(\bx_m,\by_m))\sim P^m$.
	If
	\[
		m
		\ge
		C\frac{D+\log(1/\beta)}{\eps^2}\,,
	\]
	then, with probability at least $1-\beta$, every affine halfspace $h$ on $\R^D$ satisfies
	\[
		\left|
		\widehat{\err}_{\bS}(h\circ\Phi)
		-
		\err_P(h\circ\Phi)
		\right|
		\le
		\eps\,,
	\]
	where $\err_P(h\circ\Phi):=\Pr_{(\bx,\by)\sim P}[h(\Phi(\bx))\ne\by]$.
	Consequently, on this event,
	\[
		\err_P(h_{\bS}\circ\Phi)
		\le
		\min_h\err_P(h\circ\Phi)+2\eps\,.
	\]
	The same conclusions hold for labels in $[-1,1]$ when empirical and true error are defined by
	\[
		\widehat{\err}_{\bS}(h\circ\Phi)
		:=
		\frac1m\sum_{j=1}^m\frac{1-\by_j\,h(\Phi(\bx_j))}{2}\,,
		\qquad
		\err_P(h\circ\Phi)
		:=
		\Ex_{(\bx,\by)\sim P}
		\sbra{\frac{1-\by\,h(\Phi(\bx))}{2}}\,.
	\]
\end{proposition}

\begin{proof}
	For each mapped point $\Phi(x_j)$, consider the hyperplane
	\[
		\cbra{(w,\theta)\in\R^{D+1}:w\cdot\Phi(x_j)-\theta=0}\,.
	\]
	Each face of this arrangement determines one prediction vector on the sample, including prediction vectors attained on a boundary under our convention $\sign(0)=1$.
	The arrangement has $m^{O(D)}$ faces, and a representative of each face can be found by linear programming.
	We enumerate these prediction vectors and return one making the fewest mistakes on $S$.

	The class $\{h\circ\Phi:h\text{ is an affine halfspace on }\R^D\}$ has VC dimension at most $D+1$.
	The standard VC uniform-convergence bound~\cite[Chapter~3]{KearnsVazirani:94} gives the claimed sample complexity.
	The guarantee for $h_{\bS}$ follows by comparing its sample error first with that of the best halfspace and then comparing both sample errors with their true errors.
	For labels in $[-1,1]$, the same enumeration minimizes the weighted error.
	The standard Rademacher-complexity bound for a Boolean class of VC dimension at most $D+1$ gives the same uniform-convergence guarantee.
\end{proof}

\section{Influence PCA on the Boolean Hypercube}
\label{sec:influence-pca}

In this section, we define a Boolean analogue of the Gaussian gradient second-moment matrix recalled in \Cref{eq:gaussian-PCA-matrix}. 
In Gaussian space, the top eigenvectors of this matrix identify the directions along which a smoothed function varies most significantly, and this is the central object in the influence-PCA machinery of \cite{DMN21,DKKTZ24}.  
The corresponding definition on the Boolean hypercube is not immediate: the hypercube is not rotationally invariant, and it has no intrinsic notion of differentiation in an arbitrary direction.

At a high level, the Boolean matrix will record how the degree-$1$ Fourier weight of a $\rho$-random restriction of $f:\bn\to[-1,1]$ is distributed across directions.
We establish its basic identities, relate it to the Gaussian gradient second-moment matrix, and record the spectral properties needed for our learning algorithm. 
We also give a membership-query estimator whose output is always positive semidefinite with trace at most one. 

\subsection{Definition and Basic Properties}
\label{subsec:random-restrictions}
\label{subsec:basic-identities}

For $\rho \in [0,1]$, let $\mu_{\rho,*}^{\otimes n}$ be the product distribution on $\{\pm 1, *\}^n$ where each coordinate of $\bz\sim \mu_{\rho,*}^{\otimes n}$ is drawn independently according to the following distribution: 
\[
	\bz_i = 
	\begin{cases}
		* & \text{with probability}~1-\rho \\
		+1 & \text{with probability}~\rho/2 \\	
		-1 & \text{with probability}~\rho/2 \\
	\end{cases}\,.
\]

We can now define the Boolean influence PCA matrix.

\begin{definition}[Influence PCA matrix]
\label{def:M}
	Suppose $f:\bn\to[-1,1]$ and $\rho \in [0, 1]$. 
	Sample $\bz\sim\mu_{\rho,*}^{\otimes n}$. 
	Let $S(\bz) := \{i : \bz_i = \ast\}$ be the set of free coordinates. 
	For $i \in S(\bz)$, define the degree-$1$ Fourier coefficient of the restricted function $f_{\bz}:\bits^{S(\bz)}\to[-1,1]$ at coordinate $i$ as 
	\[
		\wh{f_{\bz}}(i) := \sum_{\substack{T \subseteq [n] \\ T\cap S(\bz)=\{i\}}} \wh{f}(T) \prod_{j \in T \setminus \{i\}} \bz_j\,.
	\]
	Set $\bX_{\rho,f} \in \R^n$ by $(\bX_{\rho,f})_i = \widehat{f_{\bz}}(i)$ if $i \in S(\bz)$, and $0$
	otherwise. 
	Finally, define the \emph{influence PCA matrix} $M_\rho(f)$ as 
	\begin{equation} \label{eq:Mmat-def}
	  M_\rho(f) := \E\bigl[\bX_{\rho,f}\,\bX_{\rho,f}^\top\bigr]\,.
	\end{equation}
\end{definition}

Informally, each random restriction $\bz \sim \mu_{\rho,*}^{\otimes n}$ produces a vector $\bX_{\rho, f}$ consisting of the degree-1 Fourier coefficients of the restricted function $f_{\bz}$, padded with zeros on the restricted coordinates. 
The matrix $M_\rho(f)$ is the (uncentered) second-moment matrix of these random vectors. 
Equivalently, for every unit vector $v\in\mathbb{S}^{n-1}$, the quantity $v^\top M_\rho(f) v$ is the expected squared projection of the linear Fourier component of $f_{\bz}$ onto $v$.
We can thus view $M_\rho(f)$ as capturing a directional refinement of the expected degree-1 Fourier weight.

\begin{proposition}[Basic properties of $M_\rho(f)$]
	\label{lem:pca-basic}
	For every $f:\bits^n\to[-1,1]$ and $\rho\in[0,1]$, the following properties hold.
	\begin{enumerate}
		\renewcommand{\labelenumi}{\arabic{enumi}.}
		\item The matrix $M_\rho(f)$ is positive semidefinite.
		\item Its trace satisfies $\Tr(M_\rho(f))\leq1$.
		\item For every $i\in[n]$,
		\begin{equation}
			\label{eq:Mmat-diagonal-entry}
			M_\rho(f)_{ii}
			=
			(1-\rho)
			\sum_{A\subseteq[n]\setminus\{i\}}
			\rho^{|A|}
			\widehat f(A\cup\{i\})^2
			=
			(1-\rho)\,\Inf_i^{(\rho)}[f]\,.
		\end{equation}
		For every pair of distinct coordinates $i,j\in[n]$,
		\begin{equation}
			\label{eq:Mmat-off-diagonal-entry}
			M_\rho(f)_{ij}
			=
			(1-\rho)^2
			\sum_{A\subseteq[n]\setminus\{i,j\}}
			\rho^{|A|}
			\widehat f(A\cup\{i\})
			\widehat f(A\cup\{j\})\,.
		\end{equation}
	\end{enumerate}
\end{proposition}

The diagonal identity explains the name ``influence PCA'' as for $\rho\in(0,1)$ we have
\[
	M_\rho(f)_{ii}=\frac{1-\rho}{\rho}\,\Inf_i[\T_{\sqrt\rho}f]\,,
\]
so, up to the scaling factor $(1-\rho)/\rho$, the diagonal entries are exactly the coordinate influences of the smoothed function.

\begin{proof}
	The positive-semidefiniteness in the first property follows immediately from \Cref{eq:Mmat-def}.
	For the trace bound in the second property, Parseval's formula gives
	\begin{align*}
		\Tr(M_\rho(f)) 
		= \Ex_{\bz\sim \mu^{\otimes n}_{\rho,*}}\sbra{\|\bX_{\rho,f}\|_2^2}
		&= \Ex_{\bz\sim \mu^{\otimes n}_{\rho,*}}\sbra{\sum_{i \in S(\bz)} \widehat{f_{\bz}}(i)^2} \\
		&\leq \Ex_{\bz\sim \mu^{\otimes n}_{\rho,*}}\sbra{\sum_{T \sse S(\bz)} \wh{f_{\bz}}(T)^2} \\
		&= \Ex_{\bz\sim \mu^{\otimes n}_{\rho,*}}\sbra{\Ex_{\by\sim \bits^{S(\bz)}}\sbra{f_{\bz}(\by)^2}} \tag{Parseval's formula} \\[0.5em]
		&\leq 1\,,
	\end{align*}
	since $f$ is bounded in $[-1,1]$. 

	It remains to justify \Cref{eq:Mmat-diagonal-entry,eq:Mmat-off-diagonal-entry}. 
	When $\rho=1$, both sides of \Cref{eq:Mmat-diagonal-entry,eq:Mmat-off-diagonal-entry} vanish.
	Suppose henceforth that $\rho<1$, let $\bz\sim\mu_{\rho,*}^{\otimes n}$, and define the boldfaced random variables $\by_j$ by setting $\by_j:=\bz_j$ when $\bz_j\ne *$ and $\by_j:=0$ when $\bz_j=*$.
	The random variables $\by_1,\ldots,\by_n$ are independent and satisfy $\Ex[\by_j]=0$ and $\Ex[\by_j^2]=\rho$.
	\[
		(\bX_{\rho,f})_i
		=
		\mathbf 1\{\bz_i=*\}
		\sum_{A\subseteq[n]\setminus\{i\}}
		\widehat f(A\cup\{i\})\prod_{j\in A}\by_j\,.
	\]
	For the diagonal entry, condition on $\bz_i=*$ and use orthogonality of the monomials in the remaining independent, mean-zero random variables.
	This gives the first equality in \Cref{eq:Mmat-diagonal-entry}, including the factor $1-\rho$.
	The second equality follows from the definition of noisy influence in \Cref{subsec:fourier}; at $\rho=0$, only $A=\emptyset$ contributes, in agreement with our convention there.
	For $i\ne j$, condition on $\bz_i=\bz_j=*$.
	Terms involving $\by_i$ or $\by_j$ vanish, and orthogonality on the other coordinates leaves precisely the common index sets $A\subseteq[n]\setminus\{i,j\}$.
	Multiplying by $\Pr[\bz_i=\bz_j=*]=(1-\rho)^2$ gives \Cref{eq:Mmat-off-diagonal-entry}.
\end{proof}

\subsection{Relationship to the Gaussian Influence PCA Matrix}
\label{subsec:gradient-domination}

We next relate the Boolean influence PCA matrix $M_\rho(f)$ to its Gaussian analogue.
For a square-integrable function $f:\R^n\to\R$ and $\rho \in [0,1)$, define
\begin{equation} \label{eq:gaussian-PCA-matrix}
	\Gamma_\rho(f)
	:=
	\Ex_{\bx\sim N(0,I_n)}
	\sbra{
	\nabla \U_{\sqrt\rho} f(\bx)
	\,
	\bigl(\nabla \U_{\sqrt\rho} f(\bx)\bigr)^\top 
	}\,.
\end{equation}
This is the Gaussian influence PCA matrix used in earlier work~\cite{DMN21,DKKTZ24}.

To motivate the following lemma, we briefly explain how it will help us study $\corr_f(h)$ for a regular halfspace $h$. 
By \Cref{thm:boolean-gaussian-bridge}, the geometric Gaussian extension of $h$ is close to its multilinear extension, allowing us to approximate $\corr_f(h)$ by a corresponding correlation between functions on Gaussian space. 
The matrix $\Gamma_\rho(\wt f)$ identifies the significant Gaussian directions of the smoothed target.
The following lemma connects these directions back to the Boolean setting: every direction significant for $\Gamma_\rho(\wt f)$ is also significant for $M_\rho(f)$, up to scaling factors in $\rho$ and $1-\rho$. 

\begin{lemma}
\label{lem:gradient-domination}
	Let $\rho\in[0,1)$ and $f:\bits^n\to[-1,1]$.
	Writing $\operatorname{diag}(M)$ for the diagonal matrix with the same diagonal as $M$, we have
	\begin{equation}
		\label{eq:exact-gradient-covariance}
		\Gamma_\rho(\wt f)
		=
		\frac{\rho}{(1-\rho)^2}
		\bigl(M_\rho(f)-\rho\operatorname{diag}(M_\rho(f))\bigr)
		\preceq
		\frac{\rho}{(1-\rho)^2}M_\rho(f)\,.
	\end{equation}
\end{lemma}

\begin{proof}
	Differentiating the multilinear Hermite expansion gives
	\[
		\partial_i\U_{\sqrt\rho}\wt f(x)
		=
		\sum_{A\subseteq[n]\setminus\{i\}}
		\rho^{(|A|+1)/2}\widehat f(A\cup\{i\})\prod_{j\in A}x_j\,.
	\]
	Orthogonality of square-free monomials under the Gaussian measure therefore yields, for all $i,j$,
	\[
		\Gamma_\rho(\wt f)_{ij}
		=
		\rho\sum_{A\subseteq[n]\setminus\{i,j\}}
		\rho^{|A|}\widehat f(A\cup\{i\})\widehat f(A\cup\{j\})\,.
	\]
	Comparing with \Cref{eq:Mmat-diagonal-entry,eq:Mmat-off-diagonal-entry} proves the identity.
	The inequality follows because $\operatorname{diag}(M_\rho(f))\succeq0$.
\end{proof}

\subsection{Top Eigenspaces and Matrix Estimation}
\label{subsec:spectral-influential-subspaces}

We will use $M_\rho(f)$ to define a low-dimensional subspace of $\R^n$ over which our learning algorithms can search.
This subspace will be a top eigenspace of $M_\rho(f)$, so we next record the spectral properties and estimation guarantee that the algorithm requires.

\begin{definition}[Top eigenspace]
	\label{def:top-eigenspace}
	Let $M$ be a symmetric positive semidefinite matrix and let $\lambda>0$.
	We write $\Eigen_\lambda(M)$ for the span of the eigenvectors of $M$ with
	eigenvalue at least $\lambda$.
\end{definition}

For $M=M_\rho(f)$, the top eigenspace captures the directions in which $f$ has the most influence at smoothing scale $\rho$. 
The following lemma allows us to control the projector onto the top eigenspace of a matrix with bounded trace.  

\begin{lemma}
	\label{lem:projector-regularity}
	Let $M\succeq0$ satisfy $\Tr(M)\le1$, and let
	$E:=\Eigen_\lambda(M)$ for $\lambda>0$.
	Then
	\[
		\dim(E)\le\lambda^{-1}\,,
		\qquad
		\Pi_E\preceq\frac{M}{\lambda}\,,
	\]
	and every unit vector $v\in E^\perp$ satisfies $v^\top M v\le\lambda$.
	Moreover, if
	$M_{ii}\le t_{\mathrm{diag}}$ for every $i$, then every unit vector $u\in E$ satisfies
	\[
		\vabs{u}_\infty\le\sqrt{\frac{t_{\mathrm{diag}}}{\lambda}}\,.
	\]
\end{lemma}

\begin{proof}
	The trace bound gives $\lambda\dim(E)\le\Tr(M)\le1$.
	In an
	orthonormal eigenbasis of $M$, the projector $\Pi_E$ has eigenvalue $1$
	only where $M$ has eigenvalue at least $\lambda$, so
	$\Pi_E\preceq M/\lambda$.
	The same eigenbasis shows that a unit vector
	$v\in E^\perp$ satisfies $v^\top M v\le\lambda$.
	Finally, for a unit
	vector $u\in E$ and any $i\in[n]$,
	\[
		|u_i|^2
		\le(\Pi_E)_{ii}
		\le\frac{M_{ii}}{\lambda}
		\le\frac{t_{\mathrm{diag}}}{\lambda}\,. \qedhere
	\]
\end{proof}

The matrix $M_\rho(f)$ can be estimated directly from membership queries.
For each random restriction, two independent estimates of its degree-$1$ Fourier vector give an unbiased estimate of the outer product.
The proof uses a mean-square bound followed by projection onto the positive semidefinite trace ball, and is deferred to
\Cref{app:matrix-estimation}.

\begin{samepage}
\begin{proposition}[Membership-query matrix estimator]
	\label{thm:matrix-estimator}
	Let $f:\bits^n\to[-1,1]$, $\rho\in(0,1)$, and
	$\beta,\kappa,\delta\in(0,1/2)$, and set $\eta:=\min\{\kappa,\delta\}$.
	There is a membership-query algorithm which always returns a positive semidefinite random matrix $\widehat{\bM}$ with
	$\Tr(\widehat{\bM})\le1$ and, with probability at least $1-\beta$, satisfies
	\[
		\|\widehat{\bM}-M_\rho(f)\|_{\mathrm{op}}\le\kappa\,,
		\qquad
		\max_i|\widehat{\bM}_{ii}-M_\rho(f)_{ii}|\le\delta\,.
	\]
	Its query complexity is
	\[
		O\!\left(n\eta^{-2}\log\frac1\beta\right)\,.
	\]
	For any fixed failure probability, the construction uses
	$N=O(\eta^{-2})$ independent restrictions and two independent batches of $R=n$ uniform points per restriction.
	For general $\beta$, its running time is
	\[
		O\!\left(n^2\eta^{-2}\log\frac1\beta
		+n^2\log^2\frac1\beta+n^3\right)\,.
	\]
	There is no additional dependence on $(1-\rho)^{-1}$.
\end{proposition}
\end{samepage}

\section{Proper Agnostic Learning with Queries}
\label{sec:agnostic-learning}

We now prove \Cref{thm:main}, which we recall below for the reader's convenience.

\MainTheorem*

We briefly recall the proof strategy outlined in \Cref{subsec:overview-main}.
First, we smooth the target function $f$ and use the diagonal of $M_\rho(f)$ to identify a small set of head coordinates.
The critical-index dichotomy then allows us to regularize the tail of any candidate halfspace while preserving its head coefficients and losing little correlation with $f$. 
We next pass to Gaussian space in the tail coordinates and use a projection argument to replace the tail weights by a vector in a common low-dimensional subspace, again with little loss in correlation. 
Thus, every candidate halfspace can be replaced, with little loss in correlation with $f$, by one whose normal lies in the common low-dimensional space spanned by the head coordinates and the leading eigenspace of the tail block $M_\rho(f)[T,T]$.
We then perform empirical risk minimization over halfspaces in this space to obtain a nearly optimal hypothesis. 

\subsection{The Head Variables}
\label{subsec:global-smoothing}

Fix $\rho\in(0,1)$ and define
\begin{equation}
    \label{eq:globally-smoothed-target}
    f^{\mathrm{sm}}:=\T_{\sqrt\rho}f\,.
\end{equation}
By \Cref{lem:pca-basic},
\[
    M_\rho(f)_{ii}
    =\frac{1-\rho}{\rho}\Inf_i[f^{\mathrm{sm}}]\,.
\]
For $\alpha\in(0,1)$, define
\begin{equation}
    \label{eq:exact-global-head}
    H:=\{i:\Inf_i[f^{\mathrm{sm}}]>\rho\alpha\}
      =\{i:M_\rho(f)_{ii}>(1-\rho)\alpha\}\,.
\end{equation}
The trace bound gives $|H|<((1-\rho)\alpha)^{-1}$.
The argument below also applies to any head $H$ satisfying
\begin{equation}
    \label{eq:admissible-head-influences}
    \Inf_i[f^{\mathrm{sm}}]\le\rho\alpha
    \qquad(i\in H^c)\,.
\end{equation}
This flexibility will allow us to use an empirical head.

\subsection{Regularizing Candidate Halfspaces}
\label{subsec:global-regularization}

\begin{lemma}[Regularizing the tail]
    \label{lem:global-regular-tail}
    Let $\alpha\in(0,1)$, let $H$ satisfy \Cref{eq:admissible-head-influences}, and let
    \[
        h(x)=\sign(w_H\cdot x_H+w_{H^c}\cdot x_{H^c}-\theta)\,.
    \]
    For every $\tau,\nu\in(0,1/2)$, there is a halfspace
    \[
        h^*(x)=\sign(w_H\cdot x_H+u\cdot x_{H^c}-\theta^*)
    \]
    whose tail normal $u$ is zero or $\tau$-regular and such that
    \[
        \corr_{f^{\mathrm{sm}}}(h^*)
        \ge \corr_{f^{\mathrm{sm}}}(h)
        -C\left(\nu+\frac{\sqrt{\alpha\log(1/\nu)\Lambda_\nu}}{\tau}\right)\,.
    \]
\end{lemma}

\begin{proof}
    If $w_{H^c}=0$, take $h^*=h$.
    Otherwise order the $m:=|H^c|$ tail coordinates by decreasing absolute coefficient, and apply \Cref{thm:critical-index} with
    \[
        K:=\left\lceil C_{\mathrm{crit}}\tau^{-2}\log(1/\nu)\Lambda_\nu\right\rceil.
    \]
    Let $\ell$ be the $\tau$-critical index of this ordered tail normal.

    If $\ell\le\min\{K,m\}$, let $S$ be the first $\ell$ tail coordinates. For every $z\in\bits^S$, the halfspace obtained by pinning $x_S=z$ has the original head coefficients and a zero or $\tau$-regular tail. By \Cref{lem:coordinate-resampling}, some such assignment gives a halfspace $h^*$ with
    \[
        \corr_{f^{\mathrm{sm}}}(h^*)
        \ge\corr_{f^{\mathrm{sm}}}(h)-\sqrt{|S|\rho\alpha}
        \ge\corr_{f^{\mathrm{sm}}}(h)-\sqrt{K\alpha}.
    \]
    This also covers $m\le K$, since the empty tail is regular.

    Otherwise $K<m$. Let $S$ contain the first $K$ tail coordinates and put
    \[
        h^{\mathrm{trunc}}(x):=\sign(w_H\cdot x_H+w_S\cdot x_S-\theta).
    \]
    For each head assignment $b$, apply \Cref{thm:critical-index} with threshold $\theta-w_H\cdot b$. Averaging over $b$ shows that $h^{\mathrm{trunc}}$ disagrees with $h$ on at most a $\nu$ fraction of the cube. Since $|f^{\mathrm{sm}}|\le1$, this costs at most $2\nu$ in correlation.
    Resampling $S$ in $h^{\mathrm{trunc}}$ and using \Cref{lem:coordinate-resampling} then gives a head-only halfspace at additional cost at most $\sqrt{K\alpha}$.

    In both cases, we choose a single pinning assignment by averaging the global correlation, rather than choosing separate assignments for different head values. Thus the result has one threshold $\theta^*=\theta-w_S\cdot z$, and its head coefficients remain $w_H$.
    The total loss is at most
    \[
        2\nu+\sqrt{K\alpha}
        \le C\left(\nu+\frac{\sqrt{\alpha\log(1/\nu)\Lambda_\nu}}{\tau}\right)\,,
    \]
    as claimed. 
\end{proof}

\subsection{The Common Tail Subspace}
\label{subsec:common-subspace}

For any head $H$, write $T:=H^c$ and define 
\begin{equation}
    \label{eq:tail-principal-matrix}
    M_{\mathrm{tail}}(H):=M_\rho(f)[T,T]\,.
\end{equation}
This matrix is positive semidefinite with trace at most one. If $H$ satisfies \Cref{eq:admissible-head-influences}, then
\begin{equation}
    \label{eq:tail-matrix-diagonal}
    M_{\mathrm{tail}}(H)_{ii}
    =\frac{1-\rho}{\rho}\Inf_i[f^{\mathrm{sm}}]
    \le(1-\rho)\alpha
    \qquad(i\in T)\,.
\end{equation}

It will be convenient to collect the structural errors into $ \Delta_{\mathrm{str}} := \Delta_{\mathrm{str}}(\rho,\alpha,\tau,\nu,\lambda,\eta_E)$
\begin{equation}
    \label{eq:structural-error}
    \Delta_{\mathrm{str}}
    :=C_0\left(
    \nu+\frac{\sqrt{\alpha\log(1/\nu)\Lambda_\nu}}{\tau}
    +\sqrt{1-\rho}
    +\tau^{1/4}+\min\{1,\eta_E\}^{1/4}
    +\frac{\sqrt{\rho\lambda}}{1-\rho}
    \right)\,.
\end{equation}

\begin{theorem}
    \label{thm:axiomatic-structural}
    Let $\alpha\in(0,1)$ and $\lambda,\eta_E>0$, let $H$ satisfy \Cref{eq:admissible-head-influences}, and let $E\subseteq\R^{H^c}$ satisfy
    \[
        v^\top M_{\mathrm{tail}}(H)v\le\lambda
        \qquad(v\in E^\perp,\ \|v\|_2=1)\,.
    \]
    Suppose every vector in $E$ is $\eta_E$-regular.
    For every $\tau,\nu\in(0,1/2)$ and every halfspace $h$, there is a halfspace $g$ whose normal belongs to $\spn\{e_i:i\in H\}+E$ and such that
    \[
        \corr_f(g)\ge\corr_f(h)
        -\Delta_{\mathrm{str}}(\rho,\alpha,\tau,\nu,\lambda,\eta_E)\,.
    \]
    We identify $E$ with its embedding in $\R^n$ supported on $H^c$.
\end{theorem}

\begin{proof}
    By self-adjointness and Peres' theorem (\Cref{prop:peres}), every halfspace $a$ satisfies
    \begin{equation}
        \label{eq:global-smoothing-correlation}
        |\corr_f(a)-\corr_{f^{\mathrm{sm}}}(a)|
        =|\Ex[f(a-\T_{\sqrt\rho}a)]|
        \le C\sqrt{1-\rho}\,.
    \end{equation}
    Apply \Cref{lem:global-regular-tail} to obtain
    $h^*(x)=\sign(w_H\cdot x_H+u\cdot x_T-\theta^*)$,
    where $u$ is zero or $\tau$-regular, at correlation cost $C(\nu+\tau^{-1}\sqrt{\alpha\log(1/\nu)\Lambda_\nu})$ against $f^{\mathrm{sm}}$.

    Extend only the tail variables to Gaussian space: for $b\in\bits^H$ and $z\in\R^T$, define the function $F_b(z) : \R^T\to\R$ as 
    \[
        F_b(z):=\widetilde{f^{\mathrm{sm}}}(b,z)
              =\widetilde{(f^{\mathrm{sm}})_b}(z)\,.
    \]
    Square-free orthogonality gives $\|F_b\|_2=\|(f^{\mathrm{sm}})_b\|_2\le1$ and, for every Boolean  function $a_b : \bits^T \to \R$, 
    \begin{equation}
        \label{eq:hybrid-correlation-identity}
        \Ex_{\bx_T\sim\bits^T}[(f^{\mathrm{sm}})_b(\bx_T)\,a_b(\bx_T)]
        =\Ex_{\bz\sim N(0,I_T)}[F_b(\bz)\,\widetilde{a_b}(\bz)]\,.
    \end{equation}
    Here and below $\bb$ is uniform on $\bits^H$ and $\bz\sim N(0,I_T)$, independently. Writing
    \[
        h^*_{b,\G}(z):=\sign(w_H\cdot b+u\cdot z-\theta^*)\,,
    \]
    \Cref{thm:boolean-gaussian-bridge} and Cauchy--Schwarz imply
    \begin{equation}
        \label{eq:candidate-restriction-invariance}
        \corr_{f^{\mathrm{sm}}}(h^*)
        \le\Ex_{\bb,\bz}[F_{\bb}(\bz)h^*_{\bb,\G}(\bz)]+C\tau^{1/4}\,.
    \end{equation}
    Note that if $u=0$, then the error term $C\tau^{1/4}$ can be omitted and the above inequality is an equality.  

    We next show that the Boolean matrix controls the average outer product of the tail gradients.
    Recall that $\widetilde{f^{\mathrm{sm}}}=\U_{\sqrt\rho}\widetilde f$.
    Since this extension is multilinear, each entry of its gradient outer product has degree at most two in each coordinate.
    We can therefore replace the Boolean head coordinates by independent standard Gaussian coordinates without changing the expectation: both distributions have mean zero and variance one.
    After this replacement, the expected outer product of the tail gradients is exactly the tail block of $\Gamma_\rho(\widetilde f)$.
    Applying \Cref{lem:gradient-domination} then gives
    \begin{equation}
        \label{eq:hybrid-gradient-identity}
        \Ex_{\bb,\bz}\sbra{\nabla_TF_{\bb}(\bz)\,\nabla_TF_{\bb}(\bz)^\top}
        =\Gamma_\rho(\widetilde f)[T,T]
        \preceq\frac{\rho}{(1-\rho)^2}M_{\mathrm{tail}}(H)\,,
    \end{equation}
    where the last step is \Cref{lem:gradient-domination}.

    Decompose $u=u_E+u_\perp$ along $E\oplus E^\perp$.
    If $u_\perp\ne0$, let $v=u_\perp/\|u_\perp\|_2$, and let $P_v$ average the Gaussian coordinate along $v$, leaving the head and all orthogonal coordinates fixed.
    Since $h^*_{b,\G}$ depends on $E^\perp$ only through $v$, $P_vh^*_{b,\G}=\calA_Eh^*_{b,\G}$.
    Writing $F(b,z):=F_b(z)$ and taking all $L_2$ norms over the joint distribution of $(\bb,\bz)$, we have
    \begin{align}
        \left|\Ex_{\bb,\bz}[F_{\bb}(h^*_{\bb,\G}-\calA_Eh^*_{\bb,\G})]\right|
        &\le \|F-P_vF\|_{L_2}
        \tag{\text{Self-adjointness and Cauchy--Schwarz}}\\
        &\le \|\partial_vF\|_{L_2}
        \tag{\text{Gaussian Poincar\'{e} inequality, \Cref{lem:poincare-projection}}}\\
        &\le \frac{\sqrt{\rho\lambda}}{1-\rho}\,.
        \label{eq:averaged-projection-loss}
    \end{align}
    The final inequality follows from \Cref{eq:hybrid-gradient-identity} and $v\in E^\perp$.
    Also, note that if $u_\perp=0$, averaging changes nothing. 

    Next, note that Gaussian averaging gives a mixture over halfspaces:
    \[
        \calA_Eh^*_{b,\G}(z)
        =\Ex_{\bs\sim N(0,\|u_\perp\|_2^2)}
        \sbra{\sign(w_H\cdot b+u_E\cdot z+\bs-\theta^*)}\,.
    \]
    For each fixed $s\in\R$, define the halfspace $g_s(x):=\sign(w_H\cdot x_H+u_E\cdot x_T+s-\theta^*)$, 
    and let $g_{s,b,\G}$ be the geometric Gaussian extension of $g_s(b,\cdot)$.
    Write
    \[
        c(s):=\Ex_{\bb,\bz}\sbra{F_{\bb}(\bz)g_{s,\bb,\G}(\bz)}\,.
    \]
    The mixture representation and linearity of expectation give
    \[
        \Ex_{\bb,\bz}\sbra{F_{\bb}(\bz)\calA_Eh^*_{\bb,\G}(\bz)}
        =\Ex_{\bs}[c(\bs)]\,.
    \]
    Hence there is a single shift $s_0\in\R$ such that
    \[
        c(s_0)\ge\Ex_{\bs}[c(\bs)]
        =\Ex_{\bb,\bz}\sbra{F_{\bb}(\bz)\calA_Eh^*_{\bb,\G}(\bz)}\,.
    \]
    Set $g:=g_{s_0}$.
    This choice of $s_0$ is common to all head assignments, and the normal of $g$ belongs to $\spn\{e_i:i\in H\}+E$.
    
    It remains to relate $c(s_0)$ to the Boolean correlation of $g$ with $f^{\mathrm{sm}}$.
    Write $g_b:=g(b,\cdot)$ and $g_{b,\G}:=g_{s_0,b,\G}$.
    If $u_E\ne0$, then $u_E\in E$ is $\min\{1,\eta_E\}$-regular, so \Cref{thm:boolean-gaussian-bridge} applies to every $g_b$.
    Using \Cref{eq:hybrid-correlation-identity}, Cauchy--Schwarz, and $\|F_b\|_{L_2}\le1$, we obtain
    \begin{align*}
        \left|c(s_0)-\corr_{f^{\mathrm{sm}}}(g)\right|
        &=
        \left|\Ex_{\bb,\bz}\sbra{
            F_{\bb}(\bz)
            \bigl(g_{\bb,\G}(\bz)-\widetilde{g_{\bb}}(\bz)\bigr)
        }\right|\\
        &\le
        \Ex_{\bb}\sbra{
            \|F_{\bb}\|_{L_2}
            \|g_{\bb,\G}-\widetilde{g_{\bb}}\|_{L_2}
        }\\
        &\le C\min\{1,\eta_E\}^{1/4}\,,
    \end{align*}
    where the $L_2$ norms are with respect to $N(0,I_T)$.
    If $u_E=0$, the two extensions coincide and this error is zero.
    
    We can now put everything together: 
    \begin{align}
        \corr_{f^{\mathrm{sm}}}(h^*)
        &\le
        \Ex_{\bb,\bz}\sbra{F_{\bb}h^*_{\bb,\G}}
        +C\tau^{1/4}
        \tag{\text{\Cref{eq:candidate-restriction-invariance}}}\\
        &\le
        \Ex_{\bb,\bz}\sbra{F_{\bb}\calA_Eh^*_{\bb,\G}}
        +C\tau^{1/4}
        +\frac{\sqrt{\rho\lambda}}{1-\rho}
        \tag{\text{\Cref{eq:averaged-projection-loss}}}\\
        &\le
        c(s_0)+C\tau^{1/4}
        +\frac{\sqrt{\rho\lambda}}{1-\rho}
        \tag{\text{choice of $s_0$}}\\
        &\le
        \corr_{f^{\mathrm{sm}}}(g)
        +C\bigl(\tau^{1/4}+\min\{1,\eta_E\}^{1/4}\bigr)
        +\frac{\sqrt{\rho\lambda}}{1-\rho}\,.
        \label{eq:smoothed-structural-comparison}
    \end{align}
    
    Finally, we account for regularizing $h$ and for replacing $f$ by its smoothed version.
    Applying \Cref{eq:global-smoothing-correlation} to both $h$ and $g$ incurs a total cost of at most $2C\sqrt{1-\rho}$.
    Together with the regularization bound from \Cref{lem:global-regular-tail} and \Cref{eq:smoothed-structural-comparison}, this gives
    \begin{align*}
        \corr_f(h)-\corr_f(g)
        &\le
        \corr_{f^{\mathrm{sm}}}(h)
        -\corr_{f^{\mathrm{sm}}}(g)
        +2C\sqrt{1-\rho}\\
        &\le
        \corr_{f^{\mathrm{sm}}}(h^*)
        -\corr_{f^{\mathrm{sm}}}(g)
        +C\left(
            \nu+\frac{\sqrt{\alpha\log(1/\nu)\Lambda_\nu}}{\tau}
        \right)
        +2C\sqrt{1-\rho}\\
        &\leq\Delta_{\mathrm{str}}(\rho,\alpha,\tau,\nu,\lambda,\eta_E)\,,
    \end{align*}
    provided the universal constant $C_0$ in \Cref{eq:structural-error} is sufficiently large.
    This proves the claimed bound.
\end{proof}

\subsection{The Empirical Head and Top Eigenspace}
\label{subsec:empirical-global-subspace}

Our learning algorithm chooses the head coordinates and tail subspace using an empirical estimate of $M_\rho(f)$. 
The following lemma shows that sufficiently small estimation errors preserve the properties required by \Cref{thm:axiomatic-structural}.

\begin{lemma}
    \label{cor:empirical-structural}
    Put $t:=(1-\rho)\alpha$.
    Suppose $\widehat M\succeq0$, $\Tr(\widehat M)\le1$, and
    \[
        \|\widehat M-M_\rho(f)\|_{\mathrm{op}}\le\lambda/2,
        \qquad
        \max_i|\widehat M_{ii}-M_\rho(f)_{ii}|\le t/2\,.
    \]
    Define
    \[
        \widehat H:=\{i:\widehat M_{ii}>t/2\},\qquad
        \widehat T:=\widehat H^c,\qquad
        \widehat E(\widehat H):=\Eigen_{\lambda/2}(\widehat M[\widehat T,\widehat T])\,.
    \]
    Then $|\widehat H|\le2/t$, $\dim\widehat E(\widehat H)\le2/\lambda$, and $\widehat H$ satisfies \Cref{eq:admissible-head-influences}.
    Every unit $v\in\widehat E(\widehat H)^\perp$ satisfies
    $v^\top M_{\mathrm{tail}}(\widehat H)v\le\lambda$, and every unit $u\in\widehat E(\widehat H)$ satisfies
    \[
        \|u\|_\infty\le\sqrt{t/\lambda}\,.
    \]
\end{lemma}

\begin{proof}
    Each coordinate in $\widehat H$ contributes more than $t/2$ to the trace of $\widehat M$.
    Since $\Tr(\widehat M)\le1$, we have $|\widehat H|\le2/t$.

    For every $i\in\widehat T$, the diagonal error bound gives
    \[
        M_\rho(f)_{ii}
        \le\widehat M_{ii}+t/2
        \le t\,.
    \]
    The diagonal identity in \Cref{lem:pca-basic} therefore implies
    \[
        \Inf_i[f^{\mathrm{sm}}]
        =\frac{\rho}{1-\rho}M_\rho(f)_{ii}
        \le\frac{\rho t}{1-\rho}
        =\rho\alpha\,.
    \]
    Thus $\widehat H$ satisfies \Cref{eq:admissible-head-influences}.

    Now consider the estimated tail matrix $\widehat M[\widehat T,\widehat T]$.
    It is positive semidefinite, has trace at most one, and has all diagonal entries at most $t/2$.
    Applying \Cref{lem:projector-regularity} with eigenvalue threshold $\lambda/2$ proves the claimed dimension and regularity bounds for $\widehat E(\widehat H)$.
    The same lemma also gives
    \[
        v^\top\widehat M[\widehat T,\widehat T]v\le\lambda/2
        \qquad
        \bigl(v\in\widehat E(\widehat H)^\perp,\ \|v\|_2=1\bigr).
    \]
    To compare this with the true matrix, extend $v$ by zeros on $\widehat H$.
    The assumed operator-norm error bound then gives
    \[
        v^\top M_{\mathrm{tail}}(\widehat H)v
        \le v^\top\widehat M[\widehat T,\widehat T]v+\lambda/2
        \le\lambda\,.
    \]

    Consequently, \Cref{thm:axiomatic-structural} applies with
    $H=\widehat H$, $E=\widehat E(\widehat H)$, and $\eta_E=\sqrt{t/\lambda}$.
\end{proof}

\subsection{\texorpdfstring{Proof of \Cref{thm:main}}{Proof of the Main Theorem}}
\label{subsec:main-learning}

Let $C_0$ dominate the universal constants in \Cref{thm:axiomatic-structural} and \Cref{cor:empirical-structural},  and choose a universal constant $C_1$ sufficiently large in terms of $C_0$ and the constant in \Cref{thm:boolean-gaussian-bridge}.

We briefly recall the roles of the main parameters and specify their values in terms of $\eps$.
The parameter $\alpha$ determines the influence threshold for the head, while $\lambda$ determines the top eigenspace of the empirical tail matrix.
The parameter $\tau$ specifies the regularity of the candidate tail, and $\nu$ bounds the critical-index truncation error.
We use the following parameter setting:
\begin{equation}
	\label{eq:main-parameter-choice}
	\begin{gathered}
		q_\eps:=\frac{\eps}{C_1}\,,
		\qquad
		\delta_{\mathrm{sm}}:=q_\eps^2\,,
		\qquad
		\rho:=1-\delta_{\mathrm{sm}}\,,
		\qquad
		\lambda:=q_\eps^6\,,
		\\[0.5em]
		\tau:=q_\eps^4\,,\qquad \nu:=q_\eps\,,
		\qquad
		\alpha:=q_\eps^{12}\,.
	\end{gathered}
\end{equation}
By \Cref{cor:empirical-structural}, every unit vector $u\in\widehat E(\widehat H)$ satisfies
\[
    \|u\|_\infty
    \le\sqrt{\frac{(1-\rho)\alpha}{\lambda}}
    =q_\eps^4\,.
\]
With these parameter choices, replacing $h$ by $h^*$ in \Cref{lem:global-regular-tail} also loses at most $O(q_\eps)$ in correlation with $f^{\mathrm{sm}}$.
For $\eps\in(0,1/2)$ and a sufficiently large fixed $C_1$, we have $\rho\in(0,1)$ and $\tau,\nu\in(0,1/2)$.
Setting $\kappa=\lambda/2$, $\delta=(1-\rho)\alpha/2$, and $\beta=1/6$ also gives $\beta,\kappa,\delta\in(0,1/2)$, as required by \Cref{thm:matrix-estimator}.

If $U\in\R^{|\widehat H^c|\times d}$ has orthonormal columns spanning $\widehat E(\widehat H)$, define 
\begin{equation}
	\label{eq:global-feature-map}
	\Phi(x)
	:=
	\bigl(x_{\widehat H},U^\top x_{\widehat H^c}\bigr)
	\in\R^D\,,
	\qquad
	D:=|\widehat H|+d\,.
\end{equation}
Every affine halfspace applied after the map $\Phi$ is a proper halfspace on the original cube.
Indeed, the classifier
\[
	\sign\left(
	 w_H\cdot x_{\widehat H}
	 +w_E^\top U^\top x_{\widehat H^c}
	 -\theta
	 \right)\,,
\]
has original normal $(w_H,Uw_E)$.

\RestyleAlgo{plain}
\LinesNotNumbered
\begingroup
\DecMargin{\algomargin}
\begin{algorithm}[!t]
	\makebox[0pt][l]{%
	\fbox{%
	\begin{minipage}[ht]{\dimexpr\textwidth-2\fboxsep-2\fboxrule\relax}
		\vspace{10pt}
		\begin{center}
			\begin{minipage}[t]{0.95\linewidth}

				\ul{\textbf{Inputs}}
				\vspace{5pt}

				\begin{tabular}{ccl}
					$f$ & $:=$ & MQ access to unknown target function $f:\bits^n\to\bits$ \\
					$\eps$ & $:=$ & Target error parameter in $(0,1/2)$ \\
				\end{tabular}

				\vspace{10pt}
				\ul{\textbf{Output}}
				\vspace{5pt}

				A halfspace $g\in\calH_n$

				\vspace{10pt}
				\ul{\textbf{Parameters}}
				\vspace{5pt}

				\begin{tabular}{ccl}
					$\rho,\alpha,\tau,\nu,\lambda$ & $:=$ & Parameters in \Cref{eq:main-parameter-choice} \\
					$t_{\mathrm{head}}$ & $:=$ & $(1-\rho)\alpha$ \\
					$\kappa$ & $:=$ & $\lambda/2$ \\
					$\delta$ & $:=$ & $t_{\mathrm{head}}/2$ \\
					$\beta$ & $:=$ & $1/6$ \\
				\end{tabular}

				\vspace{10pt}
				\ul{\textbf{Algorithm}}
				\vspace{-4pt}

				\begin{enumerate}
					\item Use \Cref{thm:matrix-estimator} with parameters $(\rho,\beta,\kappa,\delta)$ to estimate the global matrix $M_\rho(f)$, obtaining the random estimate $\widehat{\bM}$.
					\item Set
					\[
						\widehat H
						:=
						\cbra{i:\widehat{\bM}_{ii}>t_{\mathrm{head}}/2}\,,
						\qquad
						\widehat T:=\widehat H^c\,,
						\qquad
						\widehat{\bM}_{\widehat T}
						:=\widehat{\bM}[\widehat T,\widehat T]\,.
					\]
					\item Let $\widehat E(\widehat H):=\Eigen_{\lambda/2}(\widehat{\bM}_{\widehat T})$, let $d:=\dim(\widehat E(\widehat H))$, and let $U\in\R^{|\widehat T|\times d}$ have orthonormal columns spanning $\widehat E(\widehat H)$.
					\item Set $D:=|\widehat H|+d$ and
					\[
						m_{\mathrm{ERM}}
						:=
						\left\lceil
						C\frac{D+1}{\eps^2}
						\right\rceil\,.
					\]
					Draw uniform points $\bx^{(1)},\ldots,\bx^{(m_{\mathrm{ERM}})}\in\bits^n$ independently of one another and of all preceding matrix-estimation randomness, query their labels, and use ERM to find an affine halfspace with parameters $(w_{H,*},w_{E,*},\theta_*)$ on the mapped points
					\[
						\Phi(\bx^{(j)})
						=
						\left(\bx^{(j)}_{\widehat H},U^\top\bx^{(j)}_{\widehat T}\right)\,.
					\]
					\item Output
					\[
						g(x)
						:=
						\sign\left(
							w_{H,*}\cdot x_{\widehat H}
							+w_{E,*}^\top U^\top x_{\widehat T}
						-\theta_*
						\right)\,.
					\]
				\end{enumerate}
			\end{minipage}
		\end{center}
		\vspace{5pt}
	\end{minipage}
	}%
	}%
	\caption{Proper agnostic learner}
	\label{alg:global-agnostic-learner}
\end{algorithm}
\endgroup

Recall that \Cref{thm:axiomatic-structural} is stated in terms of six approximation errors. 
The next corollary checks that all six approximation errors are sufficiently small for the choices in \Cref{eq:main-parameter-choice} above. 

\begin{corollary}
	\label{cor:quantitative-feature-reduction}
	Set parameters as in \Cref{eq:main-parameter-choice}. 
	Let $H\subseteq[n]$ satisfy \Cref{eq:admissible-head-influences}, and let $E(H)\subseteq\R^{H^c}$ satisfy
		\[
			v^\top M_{\mathrm{tail}}(H) v\le\lambda
			\qquad(v\in E(H)^\perp,\ \vabs{v}_2=1)\,.
		\]
	Suppose every vector in $E(H)$ is $\eta_*$-regular, where
	\[
		\eta_*:=\sqrt{\frac{(1-\rho)\alpha}{\lambda}}\,.
	\]
	Then every halfspace $h$ has a halfspace $g$ whose normal lies in $\spn\{e_i:i\in H\}+E(H)$ and such that
	\[
		\corr_f(g)\ge\corr_f(h)-\frac{\eps}{4}\,.
	\]
\end{corollary}

\begin{proof}
    The six terms in \Cref{eq:structural-error} are all $O(q_\eps)$.
    The two critical-index terms satisfy $\nu=q_\eps$ and
    \[
        \frac{\sqrt{\alpha\log(1/\nu)\Lambda_\nu}}{\tau}
        =q_\eps^2\sqrt{\log(1/q_\eps)\log(e+\log(1/q_\eps))}
        =O(q_\eps),
    \]
    since $q\sqrt{\log(1/q)\log(e+\log(1/q))}$ is bounded on $(0,1/2)$.
    We also have 
    \[
        \sqrt{1-\rho}=q_\eps,
        \qquad \tau^{1/4}=q_\eps,
        \qquad \frac{\sqrt{\rho\lambda}}{1-\rho}\le q_\eps.
    \]
    Finally, 
    \begin{equation}
        \label{eq:empirical-tail-regularity-parameter}
        \eta_*=\sqrt{\frac{q_\eps^2q_\eps^{12}}{q_\eps^6}}=q_\eps^4<1,
        \qquad \eta_*^{1/4}=q_\eps.
    \end{equation}
    Thus $\Delta_{\mathrm{str}}\le C_*q_\eps\le\eps/4$ after fixing $C_1\ge4C_*$, where $C_*$ is universal. 
    The result follows by applying \Cref{thm:axiomatic-structural}.
\end{proof}

We can now prove our main result. 

\begin{proof}[Proof of \Cref{thm:main}]
    We first prove the error guarantee.
    For the feature map $\Phi$, let
    \[
        \mathcal H_\Phi
        :=\{\ell\circ\Phi:\ell\text{ is an affine halfspace on }\R^D\}.
    \]
    Since $\Phi$ is linear, every member of $\mathcal H_\Phi$ is a halfspace on the original Boolean cube.

    Let $\mathcal E_M$ be the event that the matrix estimate satisfies the accuracy guarantees of \Cref{thm:matrix-estimator}.
    Taking $\beta=1/6$ gives $\Pr[\mathcal E_M]\ge5/6$.
    On this event, \Cref{cor:empirical-structural} shows that $\widehat H$ and $\widehat E(\widehat H)$ satisfy the hypotheses of \Cref{cor:quantitative-feature-reduction}.
    Applying that corollary to a halfspace attaining $\OPT_f$, we obtain a member of $\mathcal H_\Phi$ whose correlation with $f$ is within $\eps/4$ of optimal.
    Since
    \[
        \err_f(h)=\frac{1-\corr_f(h)}2\,,
    \]
    it follows that
    \[
        \min_{h\in\mathcal H_\Phi}\err_f(h)
        \le\OPT_f+\frac{\eps}{8}\,.
    \]

    We next analyze the ERM step.  
    Condition on any output $\widehat M$ of the matrix estimator, which fixes $\Phi$ and $\mathcal H_\Phi$.
    Let $\bS$ denote the fresh labeled sample set drawn in Step~4 of \Cref{alg:global-agnostic-learner}.
    By \Cref{prop:fixed-dimensional-erm}, a sufficiently large constant in $m_{\mathrm{ERM}}$ ensures that, with conditional probability at least $5/6$,
    \[
        \left|\widehat{\err}_{\bS}(h)-\err_f(h)\right|
        \le\frac{\eps}{8}
        \qquad\text{for every }h\in\mathcal H_\Phi\,.
    \]
    Denote this event by $\mathcal E_{\mathrm{ERM}}$.
    Since the conditional guarantee holds for every matrix estimate, we also have $\Pr[\mathcal E_{\mathrm{ERM}}]\ge5/6$ without conditioning.

    Let $h_\Phi$ minimize the true error over $\mathcal H_\Phi$, and let $g$ be the empirical minimizer returned by the algorithm.
    On $\mathcal E_M\cap\mathcal E_{\mathrm{ERM}}$,
    \begin{align*}
        \err_f(g)
        &\le\widehat{\err}_{\bS}(g)+\frac{\eps}{8}\\
        &\le\widehat{\err}_{\bS}(h_\Phi)+\frac{\eps}{8}\\
        &\le\err_f(h_\Phi)+\frac{\eps}{4}\\
        &\le\OPT_f+\frac{3\eps}{8}
        \le\OPT_f+\eps\,.
    \end{align*}
    A union bound gives $\Pr[\mathcal E_M\cap\mathcal E_{\mathrm{ERM}}]\ge2/3$, proving the error guarantee.

    We now bound the query complexity and running time.
    These bounds hold even when the matrix estimate fails its accuracy guarantees.
    Indeed, \Cref{thm:matrix-estimator} always returns a positive semidefinite matrix with trace at most one.
    Since
    \[
        t_{\mathrm{head}}=(1-\rho)\alpha=q_\eps^{14},
        \qquad
        \lambda=q_\eps^6\,,
    \]
    thresholding its diagonal entries at $t_{\mathrm{head}}/2$ and its tail eigenvalues at $\lambda/2$ gives
    \[
        |\widehat H|\le2q_\eps^{-14},
        \qquad
        \dim\widehat E(\widehat H)\le2q_\eps^{-6}.
    \]
    Consequently,
    \begin{equation}
        \label{eq:global-dimension-bound}
            D
            =|\widehat H|+\dim\widehat E(\widehat H)
            =O(\eps^{-14})\,\qquad\text{and}\qquad 
            m_{\mathrm{ERM}}
            =O((D+1)/\eps^2)
            =O(\eps^{-16})\,.
    \end{equation}

    The matrix estimator uses parameters
    \[
        \kappa=q_\eps^6/2,
        \qquad
        \delta=q_\eps^{14}/2,
        \qquad
        \beta=1/6\,,
    \]
    so $\eta:=\min\{\kappa,\delta\}=q_\eps^{14}/2$.
    By \Cref{thm:matrix-estimator}, it uses $N=O(\eps^{-28})$ independent restrictions and two independent batches of $R=n$ sample points per restriction.
    The number of matrix-estimation queries is therefore
    \[
        2NR=O(n\eps^{-28}).
    \]
    The ERM stage makes another $m_{\mathrm{ERM}}=O(\eps^{-16})$ queries, so the total query complexity remains $O(n\eps^{-28})$.

    Matrix estimation, including the final positive semidefinite projection, takes $O(n^2\eps^{-28}+n^3)$ time.
    Extracting the tail matrix and computing its eigendecomposition takes an additional $O(n^3)$ time.
    (See, for example, \cite[Section~8.3]{GolubVanLoan13}.)
    Constructing the feature vectors for the ERM sample takes
    \[
        O(m_{\mathrm{ERM}}n(D+1))
        =O(n\eps^{-30})
    \]
    time.
    Finally, \Cref{prop:fixed-dimensional-erm} bounds the ERM search time by
    \[
        m_{\mathrm{ERM}}^{O(D)}
        =2^{O(\eps^{-14}\log(1/\eps))}.
    \]
    The algorithm constructs the mapped labeled sample once and then runs ERM on it, so these costs add.
    The total running time is
    \begin{equation}
        \label{eq:query-runtime-additive}
        O(n^3+n^2\eps^{-28}+n\eps^{-30})
        +2^{O(\eps^{-14}\log(1/\eps))}\,.
    \end{equation}
    The polynomial term is at most $O(n^3\eps^{-30})$, giving the running-time bound in \Cref{thm:main}.
\end{proof}

\begin{remark}[Bounded real-valued targets]
	\label{rem:bounded-real-targets}
	The proof of \Cref{thm:main} extends to every function $f:\bits^n\to[-1,1]$.
	For a Boolean halfspace $g$, extend the distance notation by
	\[
		\err_f(g):=\frac12\Ex\sbra{|f-g|}=\frac{1-\Ex[f\,g]}2\,.
	\]
	The equality holds pointwise because $f\in[-1,1]$ and $g\in\{\pm1\}$.
	For such functions, we continue to define  $\OPT_f(\calH_n):=\min_{h\in\calH_n}\err_f(h)$.
	The structural and matrix-estimation arguments use only that $f$ is bounded by $1$.
	The final search over halfspaces uses \Cref{prop:fixed-dimensional-erm}, which also applies to labels in $[-1,1]$.
	Consequently, the same algorithm outputs a proper halfspace $g$ satisfying
	\[
		\err_f(g)\le\OPT_f(\calH_n)+\eps\,,
	\]
	with probability at least $2/3$ and with the same asymptotic query and running-time bounds.
\end{remark}

\section{\texorpdfstring{Proper Agnostic Learning with Samples}{Proper Agnostic Learning with Samples}}
\label{sec:proper-sample-learning}

We now show how to obtain the sample-based learner stated in the introduction.

\ProperSampleTheorem*

The proof has two parts. First, we use the sample-based improper agnostic learner of \cite{KKMS:08} as an oracle and boost its outputs into an explicitly evaluable bounded function which preserves correlation with \emph{every} halfspace.  
We then run the bounded-real version of our proper membership-query learner on this proxy (\Cref{rem:bounded-real-targets}) and transfer its guarantee back to the original labeled distribution. 

Fix a distribution $D$ as in \Cref{thm:proper-sample}, and write
\[
	f(x):=\Ex_D\sbra{\by\mid\bx=x}\in[-1,1]\,.
\]
For every Boolean hypothesis $h$, we have 
\begin{equation}
	\label{eq:conditional-mean-error}
	\err_D(h)
	=
	\frac{1-\corr_f(h)}2\,,
	\qquad
	\OPT_D
	=
	\frac{1-\sup_{h\in\calH_n}\corr_f(h)}2\,.
\end{equation}

Before proceeding, it is useful to see why the more direct approach of running the improper learner once and then ``properizing'' its output does not suffice.

\begin{remark}[Why one improper output does not preserve halfspace correlations]
	Suppose one instead ran the improper learner (that is, the \cite{KKMS:08} algorithm) once, obtaining a hypothesis $p$ with $\err_D(p)\le\OPT_D+\delta$, and then ran the proper membership-query learner on the Boolean target $p$ with accuracy $\eta$.
	If $h^*$ is an optimal halfspace, triangle inequalities give only
	\begin{align*}
		\err_D(g)
		&\le \err_D(p)+\err_p(g)\\
		&\le \err_D(p)+\err_p(h^*)+\eta\\
		&\le 2\err_D(p)+\OPT_D+\eta
		\le 3\OPT_D+2\delta+\eta\,.
	\end{align*}
	Thus, this argument does not establish the desired $\OPT_D+\eps$ error guarantee.   
\end{remark}

\subsection{An Improper Correlation Finder}


We use the following consequence of the polynomial-regression framework of Kalai, Klivans, Mansour, and Servedio~\cite[Theorem~5]{KKMS:08} and the Boolean-cube $L^1$ approximation bound of Feldman, Kothari, and Vondr\'ak~\cite[Corollary~15]{FKV17}.
Let $D$ be any distribution on $\bits^n\times\bits$ whose first marginal is uniform.
Given $\delta,\zeta\in(0,1)$ and independent examples $(\bx,\bb)\sim D$, an improper learner $\mathcal A_{\mathrm{imp}}$ returns, with probability at least $1-\zeta$, an explicitly evaluable Boolean hypothesis $k$ such that
\begin{equation}
	\label{eq:improper-auditor-error}
	\Prx_D[k(\bx)\ne \bb]
	\le
	\min_{h\in\calH_n}\Prx_D[h(\bx)\ne \bb]+\delta\,.
\end{equation}
Equivalently,
\begin{equation}
	\label{eq:improper-auditor-correlation}
	\Ex_D\sbra{\bb\,k(\bx)}
	\ge
	\sup_{h\in\calH_n}\Ex_D\sbra{\bb\,h(\bx)}-2\delta\,.
\end{equation}
Its sample complexity and running time are
\begin{equation}
	\label{eq:improper-auditor-complexity}
	n^{O(\delta^{-2}\log(1/\delta))}
	\poly\!\left(\frac1\delta,\log\frac1\zeta\right)\,.
\end{equation}

We now use repeated calls to the improper learner to construct an explicitly evaluable function $q:\bits^n\to[-1,1]$ whose correlation with \emph{every} halfspace is close to that of $f$.
Whenever these correlations differ by more than the desired accuracy for some halfspace, the learner provides a hypothesis that we use to update $q$.
Following the boosting approach of Gopalan, Hu, Kim, Reingold, and Wieder~\cite{GHKRW23}, we show that each update decreases the squared error $\Ex[(f-q)^2]$, which bounds the number of calls to the learner.

\begin{lemma}[Boosting while preserving halfspace correlations]
	\label{lem:halfspace-indistinguishable-proxy}
	Let $D$ be a distribution on $\bits^n\times\bits$ with uniform first marginal, and define
	\[
		f(x):=\Ex_D\sbra{\by\mid\bx=x}\in[-1,1]\,.
	\]
	For every $\gamma,\beta\in(0,1)$, using only independent labeled examples from $D$ and black-box calls to the \cite{KKMS:08} algorithm (which we will call $\mathcal A_{\mathrm{imp}}$), one can construct, with probability at least $1-\beta$, an explicitly evaluable function $q:\bits^n\to[-1,1]$ such that
	\begin{equation}
		\label{eq:proxy-indistinguishability}
		\sup_{h\in\calH_n}
		\left|\corr_f(h)-\corr_q(h)\right|
		\le\gamma\,.
	\end{equation}
	The construction makes $O(\gamma^{-2})$ calls to $\mathcal A_{\mathrm{imp}}$, each with accuracy $\gamma/16$, and uses $\poly(1/\gamma,\log(1/\beta))$ additional examples.
\end{lemma}

\begin{proof}
	We construct a sequence of functions $q_0,q_1,\ldots:\bits^n\to[-1,1]$, starting from $q_0\equiv0$.
	At each round, we either verify that $q_t$ has approximately the same correlation as $f$ with every halfspace, or use the improper learner to find an update that decreases $\Ex[(f-q_t)^2]$.
	Set
	\[
		\eta:=\frac{3\gamma}{8},
		\qquad
		R:=\left\lceil\eta^{-2}\right\rceil+1.
	\]

	We first describe round $t$.
	To apply the improper learner to the difference $f-q_t$ using only samples from $D$, draw fresh independent examples $(\bx,\by)\sim D$ and replace each label by a random sign $\bb$ satisfying
	\begin{equation}
		\label{eq:residual-label-probability}
		\Pr[\bb=1\mid\bx,\by]
		=
		\frac12+\frac{\by-q_t(\bx)}4.
	\end{equation}
	This is a valid probability because $\by,q_t(\bx)\in[-1,1]$.
	The resulting examples still have uniform first marginal, and their labels satisfy
	\[
		\Ex[\bb\mid\bx]
		=
		\frac{f(\bx)-q_t(\bx)}2.
	\]
	Run $\mathcal A_{\mathrm{imp}}$ on these examples with accuracy $\gamma/16$, obtaining a Boolean hypothesis $a_t$.
	Let
	\[
		c_t:=\Ex[(f-q_t)a_t].
	\]
	Since $\calH_n$ is closed under output negation,
	\[
		\sup_{h\in\calH_n}\Ex[(f-q_t)h]
		=
		\sup_{h\in\calH_n}|\Ex[(f-q_t)h]|.
	\]
	Thus, by \Cref{eq:improper-auditor-correlation}, whenever the learner succeeds,
	\begin{equation}
		\label{eq:proxy-residual-correlation}
		c_t
		\ge
		\sup_{h\in\calH_n}|\Ex[(f-q_t)h]|-\frac\gamma4.
	\end{equation}
	In particular, if the correlations of $f$ and $q_t$ differ substantially for some halfspace, then $a_t$ is positively correlated with $f-q_t$.

	Using fresh examples from $D$, estimate
	\[
		c_t=\Ex_D[(\by-q_t(\bx))a_t(\bx)]
	\]
	to additive accuracy $\gamma/8$, and denote the estimate by $\widehat c_t$.
	If $\widehat c_t\le\gamma/2$, stop and output $q_t$.
	Otherwise, set
	\[
		q_{t+1}(x)
		:=
		\mathrm{clip}_{[-1,1]}\bigl(q_t(x)+\eta a_t(x)\bigr).
	\]
	Here clipping replaces values below $-1$ by $-1$ and values above $1$ by $1$.
	Allow at most $R$ rounds, declaring failure if the construction has not stopped by then.

	Suppose first that every learning call and every estimate satisfies its guarantee.
	If the construction stops at round $t$, then \Cref{eq:proxy-residual-correlation} gives
	\begin{align*}
		\sup_{h\in\calH_n}|\corr_f(h)-\corr_{q_t}(h)|
		&=\sup_{h\in\calH_n}|\Ex[(f-q_t)h]|\\
		&\le c_t+\frac\gamma4\\
		&\le\widehat c_t+\frac{3\gamma}{8}\\
		&\le\frac{7\gamma}{8}
		\le\gamma.
	\end{align*}
	Thus the returned function satisfies \Cref{eq:proxy-indistinguishability}.

	To show that the construction must stop, consider the squared error
	\[
		\Phi_t:=\Ex[(f-q_t)^2].
	\]
	Whenever an update occurs,
	\[
		c_t\ge\widehat c_t-\frac\gamma8
		>\frac\gamma2-\frac\gamma8
		=\eta.
	\]
	Since $f(x)\in[-1,1]$, clipping the update cannot increase its squared distance from $f(x)$.
	Using $a_t(x)^2=1$, we therefore obtain
	\begin{align*}
		\Phi_{t+1}
		&\le\Ex[(f-q_t-\eta a_t)^2]\\
		&=\Phi_t-2\eta c_t+\eta^2\\
		&<\Phi_t-\eta^2.
	\end{align*}
	Initially $\Phi_0=\Ex[f^2]\le1$, and $\Phi_t\ge0$ for every $t$.
	Consequently, fewer than $\eta^{-2}$ updates can occur, so the construction stops within $R$ rounds.
	The returned function is explicitly evaluable: starting from zero, evaluate the stored hypotheses $a_0,\ldots,a_{t-1}$ and apply their updates and clipping operations in order.

	Finally, assign failure probability at most $\beta/(2R)$ to each learning call and each validation estimate.
	Because each uses fresh examples, these guarantees hold conditional on all preceding rounds.
	A union bound therefore gives simultaneous success with probability at least $1-\beta$.
	The number of improper-learning calls is at most $R=O(\gamma^{-2})$ on every execution. 
	Each validation observation lies in $[-2,2]$, so estimating $c_t$ to accuracy $\gamma/8$ requires $O(\gamma^{-2}\log(R/\beta))$ examples.
	The total number of validation examples is therefore $O(\gamma^{-4}\log(e/(\beta\gamma)))$, as claimed.
\end{proof}

\subsection{\texorpdfstring{Proof of~\Cref{thm:proper-sample}}{Proof of the Sample-Based Theorem}}

We can now prove our sample-based learning result. 
\begin{proof}[Proof of \Cref{thm:proper-sample}]
	We first use labeled examples to construct an explicitly evaluable function $q$ whose correlations with halfspaces approximate those of $f$.
	We then run our proper learner on $q$, answering its membership queries by evaluating $q$.

	Apply \Cref{lem:halfspace-indistinguishable-proxy} with
	\[
		\gamma:=\frac\eps4,
		\qquad
		\beta:=\frac16.
	\]
	With probability at least $5/6$, the resulting function $q:\bits^n\to[-1,1]$ satisfies
	\[
		|\corr_f(h)-\corr_q(h)|\le\gamma
		\qquad\text{for every }h\in\calH_n.
	\]
	Run the bounded-real version of the proper membership-query learner from \Cref{rem:bounded-real-targets} on $q$ with accuracy $\eta_{\mathrm{MQ}}:=\eps/4$.
	Use failure probability $1/12$ for each of its matrix-estimation and ERM steps.
	By the proof of \Cref{thm:main}, this gives success probability at least $5/6$ and changes only universal constants in the complexity bounds.
	On success, the learner returns a halfspace $g\in\calH_n$ satisfying
	\begin{equation}
		\label{eq:mq-proxy-correlation}
		\corr_q(g)
		\ge
		\sup_{h\in\calH_n}\corr_q(h)-2\eta_{\mathrm{MQ}}.
	\end{equation}
	Every query is answered exactly by evaluating $q$, so this stage requires no additional labeled examples from $D$.

	Let $h^*$ be an optimal halfspace for $D$.
	If both stages succeed, the correlation guarantee for $q$ applies to both $h^*$ and the returned halfspace $g$.
	Consequently,
	\begin{align*}
		\corr_f(g)
		&\ge \corr_q(g)-\gamma\\
		&\ge \corr_q(h^*)-2\eta_{\mathrm{MQ}}-\gamma\\
		&\ge \corr_f(h^*)-2\eta_{\mathrm{MQ}}-2\gamma.
	\end{align*}
	Converting correlations to errors using \Cref{eq:conditional-mean-error} gives
	\[
		\err_D(g)
		\le
		\OPT_D+\eta_{\mathrm{MQ}}+\gamma
		=
		\OPT_D+\frac\eps2
		\le
		\OPT_D+\eps.
	\]
	The first stage succeeds with probability at least $5/6$, and the second succeeds with probability at least $5/6$ conditional on any function $q$ returned by the first.
	A union bound therefore gives overall success probability at least $2/3$.

	We next bound the number of labeled examples and the running time.
	For convenience, write
	\[
		B_\eps:=\eps^{-2}\log(1/\eps),
		\qquad
		A_\eps:=\eps^{-14}\log(1/\eps).
	\]
	The construction of $q$ makes $O(\eps^{-2})$ calls to the improper learner, each with accuracy $\Theta(\eps)$ and failure probability $\Theta(\eps^2)$.
	By \Cref{eq:improper-auditor-complexity}, each call uses
	\[
		n^{O(B_\eps)}\poly(1/\eps)
	\]
	labeled examples and time.
	The polynomial number of calls and the additional validation examples leave the total sample bound at
	\[
		n^{O(B_\eps)}\poly(1/\eps).
	\]
	Since the proper-learning stage uses no further examples from $D$, this is also the sample complexity of the full algorithm.

	To bound the time needed to construct and evaluate $q$, recall that it is built by successive updates using $O(\eps^{-2})$ explicitly evaluable Boolean hypotheses.
	Each stored hypothesis has description size and evaluation cost at most $n^{O(B_\eps)}\poly(1/\eps)$.
	Thus, evaluating any intermediate function $q_t$, or the final function $q$, costs at most
	\[
		n^{O(B_\eps)}\poly(1/\eps)\,. 
	\]
	During construction, these evaluations are needed to generate the synthetic labels and compute the validation estimates.
	The total number of examples processed is also $n^{O(B_\eps)}\poly(1/\eps)$.
	Multiplying these bounds only changes the universal constant in the exponent, so constructing $q$ takes total time $n^{O(B_\eps)}\poly(1/\eps)$.

	At accuracy $\eps/4$, the proper learner makes $O(n\eps^{-28})$ queries to $q$, including those used to label its ERM sample.
	Combining this query count with the evaluation cost of $q$, the total time spent answering queries is
	\[
		n^{O(B_\eps)}\poly(1/\eps).
	\]
	By \Cref{eq:query-runtime-additive}, the learner's remaining computation takes
	\[
		O(n^3+n^2\eps^{-28}+n\eps^{-30})+2^{O(A_\eps)}
	\]
	time.
	The ERM search uses the mapped sample and labels already computed, so it requires no further evaluations of $q$.
	These costs therefore add, and the polynomial preprocessing cost is absorbed into $n^{O(B_\eps)}\poly(1/\eps)$.
	The total running time is
	\[
		n^{O(\eps^{-2}\log(1/\eps))}\poly(1/\eps)
		+2^{O(\eps^{-14}\log(1/\eps))},
	\]
	as claimed.
\end{proof}

\bibliographystyle{alphaabbrurl}
\bibliography{references}

\appendix
\section{Proof of~\Cref{thm:boolean-gaussian-bridge}}
\label[appendix]{app:robust-chow}

We prove the following comparison between the multilinear extension $\wt h$ and the geometric Gaussian extension $h_\G$.
We then show that the exponent in this comparison is optimal.

\BooleanGaussianBridge*

The case $w=0$ is immediate; throughout the argument below, we normalize $w$ and $\theta$ by the same positive factor so that $\|w\|_2=1$.
Throughout this section, $\bx$ is uniform on $\bits^n$ and $\bz\sim N(0,I_n)$. 
The proof uses complex exponentials, so we also allow complex-valued functions in $L^p(\Omega,\mu)$.
Their $L^p$ norm is $(\Ex_\mu[|u|^p])^{1/p}$, and the $L^2$ inner product is $\Ex_\mu[u\overline v]$.
For $u:\bits^n\to\C$, define its multilinear extension as before:
\[
    \wt u(z):=\sum_{S\subseteq[n]}\wh u(S)\prod_{i\in S}z_i,
    \qquad
    \wh u(S):=\Ex_{\bx}[u(\bx)\chi_S(\bx)].
\]
Square-free monomials are orthonormal under both the uniform Boolean measure and the standard Gaussian measure.
Consequently,
\begin{equation}
    \label{eq:complex-extension-isometry}
    \|\wt u\|_{L^2(N(0,I_n))}^2
    =\sum_{S\subseteq[n]}|\wh u(S)|^2
    =\|u\|_{L^2(\bits^n)}^2.
\end{equation}
Because multilinear extension involves only finite sums, it commutes with taking real and imaginary parts.
It also commutes with integration of a family $u_t$ whenever $t\mapsto u_t(x)$ is integrable for each Boolean input $x$.

Recall that $\beta_w:=\left(\sum_iw_i^4\right)^{1/2}$.
We will repeatedly use the bounds 
\begin{equation}
    \label{eq:bridge-beta-bounds}
    0<\beta_w\le1\,,
    \qquad
    \|w\|_\infty\le\sqrt{\beta_w}\,,
    \qquad
    \sum_i|w_i|^3\le\beta_w\le\|w\|_\infty\,.
\end{equation}
The bound on $\sum_i|w_i|^3$ follows from Cauchy--Schwarz, and the other bounds follow from $\sum_iw_i^2=1$.

\begin{figure}[!t]
    \centering
    \input{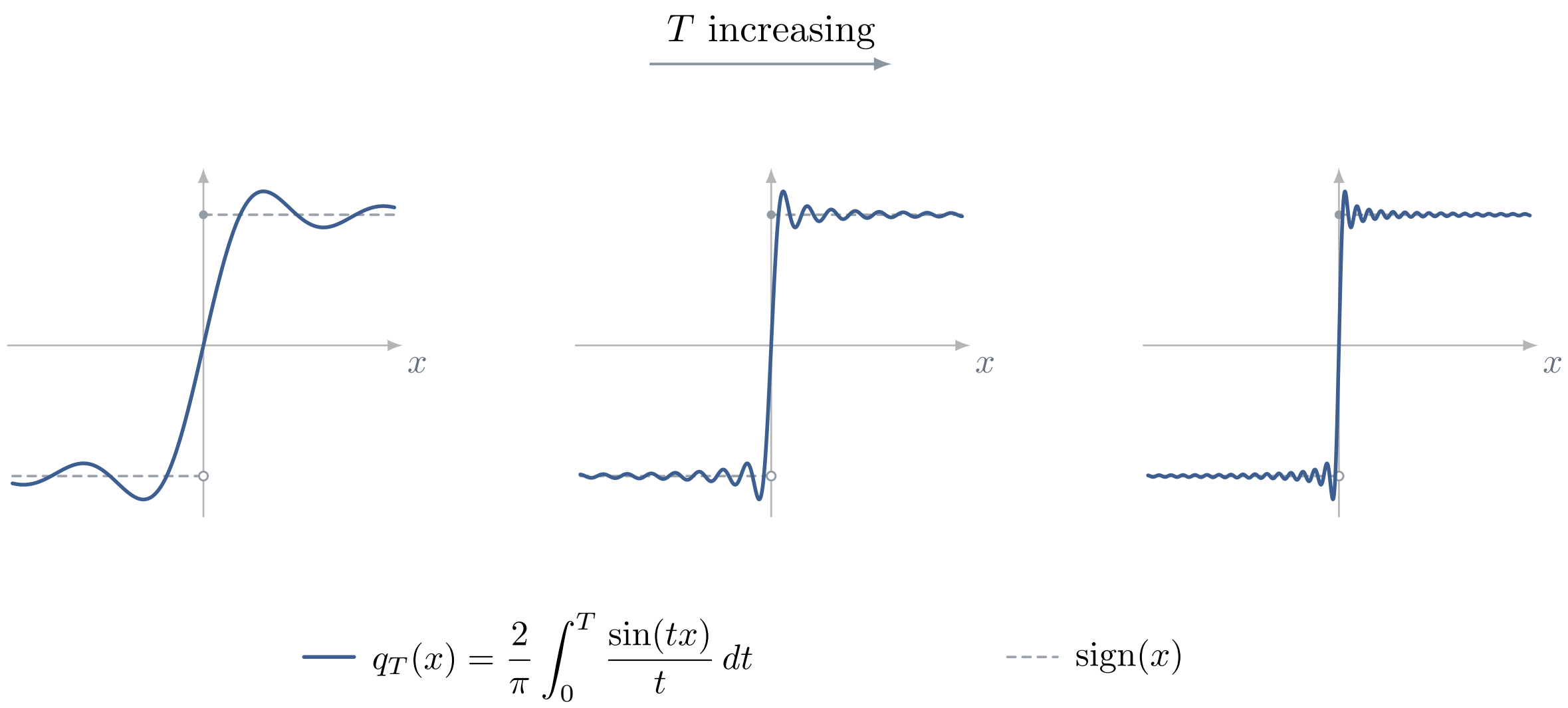}
    \vspace{0.5em}
    \caption{Approximating the $\sign$ function by an integral of sine functions.}
    \label{fig:sign-sine}
\end{figure}

At a high level, we approximate the sign function by an integral of sine functions:
\[
    \sign(v)\approx \frac2\pi\int_0^T\frac{\sin(tv)}{t}\,dt\,.
\]
As $T$ grows, this approximation becomes accurate except very close to the jump at $v=0$; see~\Cref{fig:sign-sine}. 
Regularity lets us control how often the Boolean linear form $w\cdot\bx-\theta$ lies near this jump. 
We then compare the multilinear and Gaussian extensions of each sine function appearing in the integral.
Complex numbers allow us to simplify this calculation: we write $\sin(tv)=\operatorname{Im}(e^{\mathrm{i}tv})$ and use the fact that exponentials turn sums into products.
This lets us compute the relevant expectations one coordinate at a time.
The key is to bound the average squared error of the integral directly, which gives a sharper estimate than adding the error bounds for the individual sines.
We then choose $T$ to balance the error in comparing the two extensions against the error in approximating sign. 

\subsection{Comparing Exponential Functions}

We first compare the multilinear and Gaussian extensions of the exponential function $v\mapsto e^{\mathrm{i}tv}$.
For $t\in\R$, define
\[
    P_t(x):=e^{\mathrm{i}tw\cdot x}
    \qquad(x\in\bits^n).
\]
Since
\[
    e^{\mathrm{i}tw_ix_i}
    =\cos(tw_i)+\mathrm{i}x_i\sin(tw_i)
    \qquad(x_i\in\bits),
\]
its multilinear extension has the product representation
\begin{equation}
    \label{eq:bridge-exponential-products}
    \begin{aligned}
        \wt{P_t}(z)
        &=\prod_i\bigl(\cos(tw_i)+\mathrm{i}z_i\sin(tw_i)\bigr)\,,\\
        Q_t(z)&:=e^{\mathrm{i}tw\cdot z}\,,
        \qquad
        R_t(z):=\wt{P_t}(z)-Q_t(z)\,.
    \end{aligned}
\end{equation}
Thus $R_t$ is the difference between extending $P_t$ multilinearly and evaluating the exponential directly on a Gaussian input.

Set
\[
    D(t,s):=\Ex_{\bz}\sbra{R_t(\bz)\overline{R_s(\bz)}}\,,
    \qquad
    T:=\beta_w^{-1/2}\,.
\]
The next lemma bounds these inner products.
This decay will be useful when we integrate over the frequencies.

\begin{lemma}
    \label[lemma]{lem:bridge-covariance}
    For $|t|,|s|\le T$,
    \begin{equation}
        \label{eq:bridge-covariance-bound}
        |D(t,s)|
        \le C\beta_w^2|ts|(|t|+|s|)^2e^{-(t-s)^2/4}.
    \end{equation}
\end{lemma}

\begin{proof}
    Expanding $R_t\overline{R_s}$ gives four terms, each of which factors across the Gaussian coordinates.
    To write the resulting expression compactly, define
    \[
        a(u,v):=e^{(u-v)^2/2}\cos(u-v),
        \qquad
        b(u,v):=e^{u^2/2-uv}(\cos u+v\sin u).
    \]
    For $\bZ\sim N(0,1)$, we have
    \[
        \Ex[e^{-\mathrm{i}r\bZ}]=e^{-r^2/2},
        \qquad
        \Ex[\bZ e^{-\mathrm{i}r\bZ}]
        =-\mathrm{i}r e^{-r^2/2}.
    \]
    Applying these identities coordinatewise and using $\sum_iw_i^2=1$ gives
    \begin{equation}
        \label{eq:bridge-covariance-identity}
            D(t,s)
            =e^{-(t-s)^2/2}\Bigl(
            \prod_i a(tw_i,sw_i)+1-\prod_i b(tw_i,sw_i)
            -\prod_i b(sw_i,tw_i)
            \Bigr).
    \end{equation}

    Both $a(u,v)$ and $b(u,v)$ agree with $1$ through total degree three at the origin.
    Indeed, their quadratic terms cancel, and their odd-degree terms vanish by the symmetry $(u,v)\mapsto(-u,-v)$.
    Their fourth derivatives are bounded on $|u|,|v|\le1$, so Taylor's theorem gives
    \begin{equation}
        \label{eq:bridge-factor-errors}
        |a(u,v)-1|+|b(u,v)-1|
        \le C(|u|+|v|)^4
        \qquad(|u|,|v|\le1).
    \end{equation}
    By \Cref{eq:bridge-beta-bounds}, $T\|w\|_\infty\le1$.
    Thus \Cref{eq:bridge-factor-errors} applies to every factor in \Cref{eq:bridge-covariance-identity}.

    Write each of the three products in that identity as $\prod_i(1+e_i)$.
    For each product, the factor errors satisfy
    \[
        A:=\sum_i|e_i|
        \le C\beta_w^2(|t|+|s|)^4
        \le C.
    \]
    Since $A$ is bounded by a universal constant,
    \[
        \left|\prod_i(1+e_i)-1\right|
        \le e^A-1
        \le CA.
    \]
    Substituting these bounds into \Cref{eq:bridge-covariance-identity} yields
    \begin{equation}
        \label{eq:bridge-raw-covariance}
        |D(t,s)|
        \le C\beta_w^2(|t|+|s|)^4e^{-(t-s)^2/2}.
    \end{equation}

    Note that taking $s=t$ in \Cref{eq:bridge-raw-covariance} gives
    \[
        \|R_t\|_{L^2(N(0,I_n))}^2
        =D(t,t)
        \le C\beta_w^2t^4.
    \]
    Cauchy--Schwarz therefore gives
    \[
        |D(t,s)|
        \le \|R_t\|_{L^2(N(0,I_n))}
             \|R_s\|_{L^2(N(0,I_n))}
        \le C\beta_w^2t^2s^2.
    \]
    Taking the geometric mean of this bound and \Cref{eq:bridge-raw-covariance} proves \Cref{eq:bridge-covariance-bound}.
\end{proof}

\subsection{Truncating the Sine Integral}

We next bound the error from approximating sign by a truncated sine integral.
For an arbitrary $T>0$, define
\begin{equation}
    \label{eq:bridge-sine-cutoff}
    q_T(v):=\frac2\pi\int_0^T\frac{\sin(tv)}t\,dt.
\end{equation}
The Dirichlet integral
\[
    \int_0^\infty\frac{\sin u}{u}\,du=\frac\pi2
\]
is recorded in \cite[Equation~(5.2.25)]{AbramowitzStegun:72}.
For $T|v|\ge1$, integration by parts gives
\[
    \left|\int_{T|v|}^\infty\frac{\sin u}{u}\,du\right|
    \le\frac{2}{T|v|}\,.
\]
For $T|v|\le1$, we instead use $|q_T(v)|\le2T|v|/\pi$.
Combining these estimates gives
\begin{equation}
    \label{eq:bridge-sign-error}
    |\sign(v)-q_T(v)|^2
    \le C\min\{1,(T|v|)^{-2}\}\,.
\end{equation}
At $v=0$, the minimum on the right is interpreted as $1$.
The squared error there is exactly $1$, since $\sign(0)=1$ and $q_T(0)=0$.

To turn \Cref{eq:bridge-sign-error} into an average error bound, we control the probability that the linear form lies near the threshold.
Let
\[
    F(v):=\Pr[w\cdot\bx\le v]\,,
\]
and let $\Phi$ be the standard Gaussian distribution function.
Berry--Esseen~\cite[Section~11.5]{OD14} and \Cref{eq:bridge-beta-bounds} imply
\[
    \sup_{v\in\R}|F(v)-\Phi(v)|
    \le C\sum_i|w_i|^3
    \le C\beta_w\,.
\]
For $a\le b$, write $F(a^-):=\lim_{v\uparrow a}F(v)$.
The continuity of $\Phi$ then gives
\[
    \Pr[a\le w\cdot\bx\le b]
    =F(b)-F(a^-)
    \le\Phi(b)-\Phi(a)+C\beta_w\,.
\]
Since the Gaussian density is bounded, we obtain, uniformly in $\theta\in\R$ and $r\ge0$,
\begin{equation}
    \label{eq:regular-linear-form-anticoncentration}
    \begin{aligned}
        \Pr[|w\cdot\bx-\theta|\le r]
        &\le C(r+\beta_w),\\
        \Pr[|w\cdot\bz-\theta|\le r]
        &\le Cr\,.
    \end{aligned}
\end{equation}

For any real random variable $\bV$, we have 
\[
    \Ex\bigl[\min\{1,(T|\bV|)^{-2}\}\bigr]
    =
    \int_0^1
    \Pr\bigl[|\bV|<(T\sqrt u)^{-1}\bigr]\,du.
\]
Define the truncated approximations on the two spaces by
\[
    h_T(x):=q_T(w\cdot x-\theta)
    \quad(x\in\bits^n),
    \qquad
    h_{T,\G}(z):=q_T(w\cdot z-\theta)
    \quad(z\in\R^n).
\]
Applying \Cref{eq:bridge-sign-error,eq:regular-linear-form-anticoncentration} inside the above integral, and using $\int_0^1u^{-1/2}\,du=2$, gives
\begin{equation}
    \label{eq:bridge-truncation-errors}
    \begin{aligned}
        \|h-h_T\|_{L^2(\bits^n)}^2
        &\le C(T^{-1}+\beta_w),\\
        \|h_\G-h_{T,\G}\|_{L^2(N(0,I_n))}^2
        &\le CT^{-1}\,.
    \end{aligned}
\end{equation}

\subsection{Completing the Upper Bound}

We now compare the two truncated approximations using \Cref{lem:bridge-covariance}.
We then add the truncation errors from \Cref{eq:bridge-truncation-errors}.

\begin{proof}[Proof of \Cref{thm:boolean-gaussian-bridge}]
    If $w=0$, both extensions are the same constant, so there is nothing to prove.
    Otherwise, divide $w$ and $\theta$ by the same positive factor so that $\|w\|_2=1$.
    Set
    \[
        T:=\beta_w^{-1/2}.
    \]

    We first bound the difference between the truncated approximations.
    Define
    \[
        S_{t,\theta}(z)
        :=\operatorname{Im}\bigl(e^{-\mathrm{i}t\theta}R_t(z)\bigr).
    \]
    By \Cref{eq:bridge-exponential-products,eq:bridge-sine-cutoff} and linearity of multilinear extension,
    \begin{equation}
        \label{eq:bridge-truncated-difference}
        V_T(z):=\wt{h_T}(z)-h_{T,\G}(z)
        =\frac2\pi\int_0^T\frac{S_{t,\theta}(z)}t\,dt.
    \end{equation}
    To justify passing the multilinear extension through the integral, observe that
    \[
        \frac{|\sin(t(w\cdot x-\theta))|}{t}
        \le |w\cdot x-\theta|
    \]
    for every Boolean input $x$.
    Since multilinear extension involves only finite sums, we may interchange it with the integral. 
    The Gaussian sine integral also converges in $L^2(N(0,I_n))$, since
    \[
        \left\|
            \frac{\sin(t(w\cdot z-\theta))}{t}
        \right\|_{L^2(N(0,I_n))}
        \le
        \|w\cdot z-\theta\|_{L^2(N(0,I_n))}
        =\sqrt{1+\theta^2}.
    \]

    We next express the squared norm of $V_T$ as a double integral.
    Applying \Cref{lem:bridge-covariance} with $s=t$ gives
    \[
        \frac{\|S_{t,\theta}\|_{L^2(N(0,I_n))}}t
        \le
        \frac{\|R_t\|_{L^2(N(0,I_n))}}t
        \le C\beta_w t.
    \]
    Hence, by Cauchy--Schwarz,
    \[
        \int_0^T\int_0^T
        \frac{\Ex_{\bz}[|S_{t,\theta}(\bz)S_{s,\theta}(\bz)|]}{ts}
        \,ds\,dt
        \le
        \left(
            \int_0^T
            \frac{\|S_{t,\theta}\|_{L^2(N(0,I_n))}}t\,dt
        \right)^2
        <\infty.
    \]
    Thus Fubini's theorem applies when we square \Cref{eq:bridge-truncated-difference} and take expectations.

    Since $\overline{R_s}=R_{-s}$, we can bound the inner products of the $S_{t,\theta}$ using \Cref{lem:bridge-covariance} at both $(t,s)$ and $(t,-s)$.
    More explicitly, the identity
    \[
        \Ex[\operatorname{Im}A\operatorname{Im}B]
        =
        \frac12\operatorname{Re}
        \bigl(\Ex[A\overline B]-\Ex[AB]\bigr)
    \]
    gives, for $0<t,s\le T$,
    \[
        \frac{
            \left|\Ex_{\bz}[S_{t,\theta}(\bz)S_{s,\theta}(\bz)]\right|
        }{ts}
        \le
        C\beta_w^2(t+s)^2
        \left(
            e^{-(t-s)^2/4}+e^{-(t+s)^2/4}
        \right).
    \]
    This bound is uniform in $\theta$, since the factors involving the threshold have modulus one.

    For $t,s\ge0$, we have $e^{-(t+s)^2/4}\le e^{-(t-s)^2/4}$.
    Also, $(t+s)^2\le4T^2$ on the domain of integration.
    Therefore,
    \begin{align}
        \label{eq:bridge-integrated-covariance}
        \|V_T\|_{L^2(N(0,I_n))}^2
        &\le
        C\beta_w^2
        \int_0^T\int_0^T
        (t+s)^2e^{-(t-s)^2/4}\,ds\,dt
        \notag\\
        &\le
        C\beta_w^2T^2
        \int_0^T\int_{\R}
        e^{-(t-s)^2/4}\,ds\,dt
        \notag\\
        &\le C\beta_w^2T^3.
    \end{align}
    For each fixed $t$, the integral over $s$ is bounded by a universal constant.
    Because the integrand is small when $t$ and $s$ are far apart, the double integral is $O(T)$ instead of $O(T^2)$. 

    Finally, apply the triangle inequality,  \Cref{eq:complex-extension-isometry}, and the truncation bounds \Cref{eq:bridge-truncation-errors}:
    \begin{align*}
        \|\wt h-h_\G\|_{L^2(N(0,I_n))}
        &\le
        \|h-h_T\|_{L^2(\bits^n)}
        +\|V_T\|_{L^2(N(0,I_n))}\\
        &\qquad
        +\|h_{T,\G}-h_\G\|_{L^2(N(0,I_n))}\\
        &\le
        C\left(
            \sqrt{T^{-1}+\beta_w}
            +\beta_wT^{3/2}
            +T^{-1/2}
        \right)\\
        &\le C\beta_w^{1/4}.
    \end{align*}
    The last inequality uses $T=\beta_w^{-1/2}$ and $0<\beta_w\le1$.
    Squaring gives the claimed bound $C\beta_w^{1/2}$.
    If $w$ is $\tau$-regular, then \Cref{eq:bridge-beta-bounds} gives $\beta_w\le\|w\|_\infty\le\tau$, proving the bound $C\tau^{1/2}$ as well.
\end{proof}

\subsection{Optimality of the Exponent}

We now give a matching lower bound using the majority function.
In fact, its geometric Gaussian extension cannot be approximated more closely by any multilinear polynomial, including its own multilinear extension.

\begin{proposition}[Matching lower bound]
    \label{prop:boolean-gaussian-bridge-sharpness}
    For every sufficiently large odd $n$, let $w_i=n^{-1/2}$, let $\theta=0$, and let
    \[
        h(x)=\MAJ_n(x)
        :=\sign\!\left(n^{-1/2}\sum_{i=1}^n x_i\right).
    \]
    There is a universal constant $c>0$ such that every multilinear polynomial $p:\R^n\to\R$ satisfies
    \[
        \|p-h_\G\|_{L^2(N(0,I_n))}
        \ge cn^{-1/8}.
    \]
    In particular,
    \[
        \|\wt h-h_\G\|_{L^2(N(0,I_n))}
        \ge c\tau^{1/4}
    \]
    for $\tau=n^{-1/2}$, and $\beta_w=\tau$.
\end{proposition}

\begin{proof}
    We compare the Hermite expansion of $h_\G$ with the terms that can occur in a multilinear polynomial.
    Let $\mathrm{He}_k$ be the probabilists' Hermite polynomials, and put
    \[
        H_k:=\frac{\mathrm{He}_k}{\sqrt{k!}}.
    \]
    The polynomials $(H_k)$ form an orthonormal basis for $L^2(N(0,1))$.
    For a multi-index $\alpha\in\{0,1,2,\ldots\}^n$, write
    \[
        |\alpha|:=\sum_i\alpha_i,
        \qquad
        \alpha!:=\prod_i\alpha_i!,
        \qquad
        w^\alpha:=\prod_iw_i^{\alpha_i},
        \qquad
        H_\alpha(z):=\prod_iH_{\alpha_i}(z_i).
    \]
    The $H_\alpha$ form an orthonormal basis for $L^2(N(0,I_n))$~\cite[Proposition~11.33]{OD14}.
    Since $H_0=1$ and $H_1(z)=z$, the multilinear polynomials are exactly the span of those $H_\alpha$ for which every $\alpha_i$ belongs to $\{0,1\}$.
    Thus, the squared $L^2$ distance from $h_\G$ to any multilinear polynomial is at least the sum of the squared Hermite coefficients of $h_\G$ over all $\alpha$ with some $\alpha_i\ge2$.

    We first compute the one-dimensional Hermite coefficients of the sign function.
    Write
    \[
        \sign(v)=\sum_{k\ge0}a_kH_k(v)
    \]
    in $L^2(N(0,1))$.
    The even coefficients vanish since it is an odd function. 
    Let $\varphi(v):=(2\pi)^{-1/2}e^{-v^2/2}$.
    By Rodrigues' formula~\cite[Eq.~(11.10)]{OD14}, 
    \[
        (\varphi\mathrm{He}_{k-1})'
        =-\varphi\mathrm{He}_k
    \]
    and integrating by parts gives
    \[
        a_{2m+1}
        =
        \frac{2\varphi(0)\mathrm{He}_{2m}(0)}
             {\sqrt{(2m+1)!}},
        \qquad
        \mathrm{He}_{2m}(0)
        =
        (-1)^m\frac{(2m)!}{2^m m!}.
    \]
    Consequently,
    \begin{equation}
        \label{eq:bridge-sign-hermite-coefficients}
        a_{2m+1}^2
        =
        \frac2\pi
        \frac{\binom{2m}{m}}{4^m(2m+1)}
        =
        \Theta((2m+1)^{-3/2}),
    \end{equation}
    where the final estimate follows from the standard bound on the central binomial coefficient.

    We next determine how much of each term $H_k(w\cdot z)$ is multilinear in $z$.
    Comparing coefficients in
    \[
        e^{tw\cdot z-t^2/2}
        =
        \prod_i e^{tw_i z_i-t^2w_i^2/2}
    \]
    gives, for every unit vector $w$,
    \begin{equation}
        \label{eq:bridge-hermite-addition}
        H_k(w\cdot z)
        =
        \sum_{|\alpha|=k}
        \sqrt{\frac{k!}{\alpha!}}\,w^\alpha H_\alpha(z).
    \end{equation}
    Since $w\cdot\bz$ is standard Gaussian, $H_k(w\cdot z)$ has squared $L^2$ norm one.
    For $w_i=n^{-1/2}$ and $k\le n$, its squared norm on the square-free indices is
    \[
        \binom nk\frac{k!}{n^k}
        =
        \frac{(n)_k}{n^k},
        \qquad
        (n)_k:=n(n-1)\cdots(n-k+1).
    \]
    Thus the degree-$k$ component of $h_\G$ has squared norm
    \[
        a_k^2\left(1-\frac{(n)_k}{n^k}\right)
    \]
    outside the space of multilinear polynomials.

    For degrees of order $\sqrt n$, the factor in parentheses is bounded below by a positive constant.
    Indeed, for $n\ge25$ and $\sqrt n\le k\le2\sqrt n$,
    \[
        \frac{(n)_k}{n^k}
        =
        \prod_{j=0}^{k-1}\left(1-\frac jn\right)
        \le
        \exp\!\left(-\frac{k(k-1)}{2n}\right)
        \le e^{-1/4}.
    \]
    Orthogonality and \Cref{eq:bridge-sign-hermite-coefficients} therefore imply that every multilinear polynomial $p$ satisfies
    \[
        \|p-h_\G\|_{L^2(N(0,I_n))}^2
        \ge
        \sum_{\substack{\sqrt n\le k\le2\sqrt n\\ k\ \mathrm{odd}}}
        a_k^2\left(1-\frac{(n)_k}{n^k}\right)
        \ge cn^{-1/4}.
    \]
    The last sum contains $\Omega(\sqrt n)$ terms, each at least $cn^{-3/4}$.
    Taking square roots proves the claim, including the case $p=\wt{\MAJ_n}$.
\end{proof}

Thus the exponent $1/4$ in \Cref{thm:boolean-gaussian-bridge} is optimal, even if we allow an arbitrary multilinear polynomial in place of $\wt h$.
\section{Estimating the Influence PCA Matrix}
\label[appendix]{app:matrix-estimation}

Recall that $M_\rho(f)$ is the expected outer product of the degree-$1$ Fourier vector of a random restriction.
We estimate this expectation by sampling restrictions and using membership queries to estimate their Fourier vectors.
For each restriction, we form the outer product of two independent vector estimates.
This gives an unbiased matrix estimate even when the individual vector estimates are noisy.
We then average over restrictions and use a final projection to enforce the positive-semidefiniteness and trace requirements.

We measure the estimation error in Frobenius norm,
\[
    \|A\|_F:=\left(\sum_{i,j}A_{ij}^2\right)^{1/2}\,.
\]
Since $\|A\|_{\mathrm{op}}\le\|A\|_F$ and $|A_{ij}|\le\|A\|_F$ for every entry, one Frobenius-norm bound will give both accuracy guarantees in \Cref{thm:matrix-estimator}.

\begin{proof}[Proof of \Cref{thm:matrix-estimator}]
Write $M:=M_\rho(f)$ and recall that $\eta=\min\{\kappa,\delta\}$.
Draw a restriction $\bZ$ from $\mu_{\rho,*}^{\otimes n}$, let $\bS$ be its free coordinates, and let $\bX$ be its degree-$1$ Fourier vector, padded with zeros outside $\bS$.
Parseval and the definition of $M$ give
\[
    \|\bX\|_2\le1,
    \qquad
    \Ex[\bX\bX^\top]=M\,.
\]
Conditional on $\bZ$, draw a uniform point $\bY\in\bits^{\bS}$ and define
\[
    \bW_i:=
    \begin{cases}
        f_{\bZ}(\bY)\bY_i,&i\in\bS\\
        0,&i\notin\bS
    \end{cases}
    \,.
\]
One membership query gives $f_{\bZ}(\bY)$ and hence all coordinates of $\bW$.
Moreover,
\[
    \Ex[\bW\mid\bZ]=\bX,
    \qquad
    \Ex[\|\bW\|_2^2\mid\bZ]\le|\bS|\le n\,.
\]
When $\bS$ is empty, both $\bX$ and $\bW$ are zero.

For each of $N$ independent restrictions $\bZ^{(t)}$, form two sample means $\ba_t,\bb_t$, each using $R=n$ independent copies of $\bW$.
The two batches are independent conditional on the restriction, and all samples belonging to different restrictions are independent.
The second moment of each sample mean satisfies
\[
    \Ex[\|\ba_t\|_2^2\mid\bZ^{(t)}]
    =
    \|\bX^{(t)}\|_2^2
    +
    \frac1R
    \Ex[\|\bW-\bX^{(t)}\|_2^2\mid\bZ^{(t)}]
    \le1+\frac nR
    =2\,,
\]
and the same bound holds for $\bb_t$.

Define the symmetric matrices
\[
    \bG_t:=
    \frac{\ba_t\bb_t^\top+\bb_t\ba_t^\top}{2},
    \qquad
    \bA:=\frac1N\sum_{t=1}^N\bG_t\,.
\]
Conditional independence of the two sample means gives
\[
    \Ex[\bG_t\mid\bZ^{(t)}]
    =
    \bX^{(t)}\bX^{(t)\top},
    \qquad
    \Ex[\bA]=M\,.
\]
It also gives a bound on the second moment:
\[
    \|\bG_t\|_F
    \le
    \|\ba_t\|_2\|\bb_t\|_2,
    \qquad
    \Ex[\|\bG_t\|_F^2\mid\bZ^{(t)}]\le4\,.
\]
Since the matrices $\bG_t$ are independent and identically distributed,
\begin{equation}
    \label{eq:matrix-mean-error}
    \Ex\|\bA-M\|_F^2
    =
    \frac1N\Ex\|\bG_1-M\|_F^2
    \le
    \frac1N\Ex\|\bG_1\|_F^2
    \le
    \frac4N\,.
\end{equation}
Taking $N=\lceil C\eta^{-2}\rceil$ for a sufficiently large universal constant $C$, Markov's inequality therefore gives
\[
    \Pr[\|\bA-M\|_F>\eta/3]\le\frac16\,.
\]

We next amplify the success probability.
Perform $B=O(\log(1/\beta))$ independent runs, producing estimates $\bA_1,\ldots,\bA_B$.
For each estimate $\bA_i$, count the indices $j$ for which $\|\bA_i-\bA_j\|_F\le2\eta/3$, including $j=i$.
Select any $\bA_i$ for which this count exceeds $B/2$.
If no such estimate exists, select the zero matrix. 
We will call the selected matrix $\bA_{\mathrm{sel}}$. 

Call a run ``good'' if its estimate is within Frobenius distance $\eta/3$ of $M$.
Hoeffding's inequality implies that a strict majority of runs are good with probability at least $1-\beta$.
On this event, any two good estimates are within distance $2\eta/3$ of each other, so every good estimate qualifies for selection.
Conversely, every qualifying estimate is within distance $2\eta/3$ of at least one good estimate, and hence within distance $\eta$ of $M$.
Thus, with probability at least $1-\beta$,
\[
    \|\bA_{\mathrm{sel}}-M\|_F\le\eta\,.
\]
For fixed confidence, a single run with a suitable absolute constant in $N$ also suffices.

The selected matrix is symmetric, but it need not be positive semidefinite or have trace at most one.
To enforce these requirements, project $\bA_{\mathrm{sel}}$ in Frobenius norm onto the closed convex set
\[
    \mathcal C:=
    \{K\in\R^{n\times n}:K=K^\top,\ K\succeq0,\ \Tr(K)\le1\}\,,
\]
and return $\widehat{\bM}:=\Pi_{\mathcal C}(\bA_{\mathrm{sel}})$.
Since $M\in\mathcal C$, this projection cannot increase the distance to $M$:
\[
    \|\widehat{\bM}-M\|_F
    \le
    \|\bA_{\mathrm{sel}}-M\|_F\,.
\]
Indeed, the projection inequality
\[
    \langle
        \bA_{\mathrm{sel}}-\widehat{\bM},
        M-\widehat{\bM}
    \rangle_F
    \le0
\]
gives this bound by expanding $\|\bA_{\mathrm{sel}}-M\|_F^2$.

The output therefore always satisfies the positive-semidefiniteness and trace requirements.
On the success event, it also satisfies
\[
    \|\widehat{\bM}-M\|_{\mathrm{op}}\le\eta\le\kappa,
    \qquad
    \max_i|\widehat{\bM}_{ii}-M_{ii}|\le\eta\le\delta\,.
\]

Finally, we account for the queries and running time.
Each run uses $2NR=O(n\eta^{-2})$ membership queries.
Accumulating the sample means takes $O(nNR)$ time, and forming the symmetrized outer products takes $O(n^2N)$ time, both of which are $O(n^2\eta^{-2})$.
Comparing all pairs of estimates takes $O(n^2B^2)$ time, and the final projection takes $O(n^3)$ time.
The total query complexity is therefore
\[
    O\!\left(n\eta^{-2}\log\frac1\beta\right)\,,
\]
and the total running time is
\[
    O\!\left(
        n^2\eta^{-2}\log\frac1\beta
        +n^2\log^2\frac1\beta
        +n^3
    \right)\,,
\]
as claimed.
\end{proof}

\end{document}